\documentclass[12pt,nofootinbib,notitlepage,superscriptaddress,singlecolumn]{revtex4-1}

\pdfoutput=1
\usepackage{amsmath}
\usepackage{amsfonts}
\usepackage{amssymb}
\usepackage[utf8]{inputenc} 		
\usepackage[ngerman,english]{babel} 
\usepackage[T1]{fontenc}
\usepackage{graphicx}
\usepackage{xcolor}
\usepackage[bookmarks,plainpages=false,pdfpagelabels]{hyperref}
\usepackage{natbib}
\usepackage{bbold}

\usepackage[caption=false]{subfig}

\newcommand{\D}{\mathrm{d}}
\newcommand{\eV}{\,\mathrm{eV}}
\newcommand{\kpc}{\,\mathrm{kpc}}
\newcommand{\Mpc}{\,\mathrm{Mpc}}
\newcommand{\tr}[1]{\mathrm{tr}\left(#1\right)}

\renewcommand{\Re}{\mathrm{Re}}

\begin{document}

\centerline{CP$^3$-Origins-2018-33 DNRF90}

\vspace{1truecm}

\title{Long Range Effects in Gravity Theories with Vainshtein Screening}

\author{Moritz Platscher}
\email{moritz.platscher@mpi-hd.mpg.de}
\affiliation{Max-Planck-Institut für Kernphysik,\\
					Saupfercheckweg 1, 69117 Heidelberg, Germany}
\author{Juri Smirnov}
\email{smirnov@cp3.sdu.dk}
\affiliation{$\text{CP}^3$-Origins $\&$ DIAS, University of Southern Denmark, Campusvej 55, 5230 Odense, Denmark}	
			
\author{Sven Meyer}
\email{sven.meyer@uni-heidelberg.de}
\affiliation{Universität Heidelberg, Zentrum für Astronomie,\\
					Institut für Theoretische Astrophysik,\\	
					Philosophenweg 12, 69120 Heidelberg, Germany}				 
\author{Matthias Bartelmann}
\email{bartelmann@uni-heidelberg.de}			
\affiliation{Universität Heidelberg, Zentrum für Astronomie,\\
					Institut für Theoretische Astrophysik,\\	
					Philosophenweg 12, 69120 Heidelberg, Germany}	 	
					
\begin{abstract}

In this paper we study long range modifications of gravity in the consistent framework of bigravity, which introduces a second massive spin-2 field and allows to continuously interpolate between the regime of General Relativity (mediated by a massless spin-2 field) and massive gravity (mediated by a massive spin-2 field). In particular we derive for the first time the equations for light deflection in this framework and study the effect on the lensing potential of galaxy clusters. By comparison of kinematic and lensing mass reconstructions, stringent bounds can be set on the parameter space of the new spin-2 fields. Furthermore, we investigate galactic rotation curves and the effect on the observable dark matter abundance within this framework.

\end{abstract}

\maketitle

\section{Introduction}

There is overwhelming observational evidence that the largest fractions of the matter-energy content of our universe are not properly understood in terms of their physical properties and are thus commonly termed dark. The observational evidence is entirely based on phenomena on large scales measured via gravitational effects. However, the strongest attempts to address such phenomena as dark matter (DM) are focused on particle physics models. This has to do, on the one hand, with the fact that our understanding of gravity, which roots in the theory of Einstein's General Relativity (GR), has withstood so many experimental tests across a vast range of length scales. On the other hand, while it is straightforward to modify gravity at short scales, it turns out to be conceptually very challenging  to modify GR at large scales without running into inconsistencies, such as ghost degrees of freedom, or contradictions with experiments on shorter scales. However, it would be desirable to test if gravity can indeed be modified at large scales, and what this would imply for such observables as the mass density of DM. This investigation is crucial, since our attempts to test the above mentioned particle physics models are guided strongly by assumptions about the local DM density, which on the other hand is calculated under the assumption of pure Einstein GR.

A sufficiently wide framework of consistent extensions of GR, has in fact recently been discovered, based on multi-metric gravity~\cite{Hinterbichler:2012cn}. For the sake of concreteness, we will focus on the bimetric case in this work, where two tensor fields are present. The theory is constructed such that one tensor mode remains massless while the other obtains a non-vanishing mass, comprising in total seven degrees of freedom. A priori, such a theory contains too many degrees of freedom and, related to that, massive gravity is often plagued by the Boulware-Deser ghost~\cite{Boulware:1973my}. This was a long standing issue, beginning with the proposal of a linear spin-2 theory in the first half of the 20\textsuperscript{th} century by Fierz and Pauli~\cite{Fierz:1939ix, Fierz:1939zz}. Only recently it has been proven~\cite{deRham:2010ik,deRham:2010kj,deRham:2011rn,Hassan:2011hr,Hassan:2011vm,Hassan:2011tf,Comelli:2012vz,Deffayet:2012nr,Deffayet:2012zc} that the construction of massive gravity using a second reference metric removes these pathological degrees of freedom and is free of the Boulware-Deser ghost. This framework, due to de Rham, Gabadadze and Tolley (dRGT), was later extended to a bimetric theory in Refs.~\cite{Hassan:2011zd,Hassan:2011ea}, where both tensors are dynamical. An important feature of theories of massive gravity is that, at small distances, GR is restored by non-linear effects resolving the long-standing van Dam-Veltman-Zakharov discontinuity found when the graviton mass goes to zero.~\cite{vanDam:1970vg,Zakharov:1970cc} This phenomenon was conjectured by Vainshtein already in the 1970s~\cite{Vainshtein:1972sx} and has been demonstrated explicitly for bimetric and dRGT massive gravity~\cite{Babichev:2009us,Babichev:2009ee,Babichev:2009jt,Babichev:2010jd,Babichev:2013pfa,Gruzinov:2011mm,Volkov:2012wp,Platscher:2016adw}. 

Let us first give a brief overview of existing studies considering similar effective, non-relativistic potentials as ours. Following a pioneering work~\cite{Sanders:1984aa}, most follow-up studies of modified gravity models that induce mixed Yukawa-Newton potentials, tried to argue that DM is obsolete. This was done either by means of a proof of principle studying individual examples~\cite{Drummond:2001rj,Stabile:2013jon}, or larger data sets~\cite{Moffat:2013sja,Rahvar:2014yta}, some even including galaxy clusetrs as we do~\cite{Capozziello:2008ny, Mota:2011iw}. While these studies argued against the existence of DM, others were less biased allowing a for a DM component~\cite{Piazza:2003ri,Stabile:2011zp}. And yet others quantified the observational bias on galaxy haloes that would be induced by assuming instead of the Yukawa force, Newtonian gravity~\cite{Cardone:2011ze}. A more recent idea, including for the first time the phenomenological impact of the Vainshtein screening, was made in Ref.~\cite{Enander:2015kda}. There, the authors have derived constraints by demanding that the galactic scale is contained entirely in the Vainshtein regime, such that the GR predictions apply. Also, an attempt was made recently in studying the long-range effects of dRGT massive gravity on galaxy dynamics~\cite{Panpanich:2018cxo}; however, only in a specific regime, and once again neglecting the Vainshtein screening. Similarly, in Ref.~\cite{deAlmeida:2018kwq} the authors explored generic Yukawa-type modifications of the gravitational potential on galaxy rotation curves, also disregarding the Vainshtein effect. See also \cite{2018arXiv180701725H} for a recent review. 

Our work, in contrast, focusses on bimetric gravity while simultaneously trying to be as unprejudiced as possible. To be more specific, we put the model to test at galactic and extra-galactic scales, attempting to account for the Vainshtein screening mechanism. We quantify how large a deviation from GR one actually finds, how much of it is tolerable and maybe even favourable in the sense of improving the fits to the data. In addition, we study not one scale of physical systems, namely galaxies, but also take into account galaxy clusters, and attempt to implement the Vainshtein mechanism into our phenomenological survey. This work is only meant as a pilot study of some individual measurements, and is by no means a full-fledged survey of galactic and extra-galactic effects. However, we believe that the results are encouraging for further, more detailed studies. 

The remainder of this manuscript is structured as follows. In Sec.~\ref{sec:framework} we introduce the framework and derive the induced non-relativistic gravitational potential and the lensing potential in bigravity. In Sec.~\ref{sec:clusters} we use these results to obtain constraints on the massive spin-2 field from cluster X-ray and weak lensing observations. In Sec.~\ref{sec:rotCurves} we study galaxy rotation curves and the impact on the expected DM abundance. In Sec.~\ref{sec:scaling} a general scaling argument for systems with a Vainshtein mechanism is discussed. Finally, we summarise and discuss our results in Sec.~\ref{sec:discussion}.

\section{The Bimetric Framework} 
\label{sec:framework}
We study a variant of the dRGT theory of massive gravity, which gives a simple explanation for late-time acceleration in terms of the graviton mass~\cite{Volkov:2011an,vonStrauss:2011mq,Comelli:2011zm,Akrami:2012vf,Konnig:2013gxa,Comelli:2014bqa,DeFelice:2014nja,Nersisyan:2015oha,Akrami:2015qga}. However, it has been found that phenomenologically viable, cosmological solutions are unstable and therefore some modifications are required~\cite{deRham:2010tw,DAmico:2011eto,Gumrukcuoglu:2011ew,Gumrukcuoglu:2011zh,Vakili:2012tm,DeFelice:2012mx,Fasiello:2012rw,DeFelice:2013awa}. Arguably, the simplest modification is to render the background metric dynamical itself and thereby obtain what is called bimetric gravity, or simply bigravity~\cite{Hassan:2011zd}. In fact, one could argue that bigravity is the more fundamental framework, as it arises naturally from extra-dimensional models, e.g.~via dimensional deconstruction~\cite{ArkaniHamed:2001nc,deRham:2013awa}. Following such a procedure, one is led to matter coupling diagonally to the two tensors, i.e.~visible and hidden matter sectors each have their own metric tensor. While more exotic settings display some interesting cosmological effects~\cite{Gumrukcuoglu:2014xba,Comelli:2015pua,Heisenberg:2016dkj,Gao:2016dwl,Bouhmadi-Lopez:2016fgd}, we will disregard such a hidden matter sector, and focus on the visible matter coupling only to one tensor --~including however a DM component~-- as other couplings typically lead to inconsistencies~\cite{deRham:2014naa,Hassan:2014gta,Schmidt-May:2014xla,deRham:2014fha,Yamashita:2014fga}.

The action for the two tensor fields $g$ and $f$ is given by
\begin{equation}\label{eq:action}
  \begin{gathered}
    S_\mathrm{bi} =   \int \mathrm{d}^4x \left\lbrace\frac{M_g^2}{2} \sqrt{-\det g}\, R_g + \frac{M_f^2}{2}  \sqrt{-\det f}\, R_f +
     m^2 M_\text{eff}^2 \sqrt{-\det g}\, \sum_{n=0}^4 \beta_n e_n(\mathbb{X}) \right. + \\ 
   	 \hspace{10.7cm}\left. + \sqrt{-\det g}\, \mathcal{L}_\textrm{matter} \right\rbrace.
  \end{gathered}
\end{equation}
Here, $M_g$ is the Planck scale for the physical $g$ metric, $M_f$ the Planck scale for the $f$ metric, $M_\text{eff}^2 \equiv \left( \frac{1}{M_g^2}+  \frac{1}{M_f^2}\right)^{-1}$,  and $\mathbb{X}$ is defined via the relation $\mathbb{X}^\mu_\alpha \mathbb{X}^\alpha_\nu = g^{\mu\alpha}f_{\alpha\nu}$. It is a rank-2 tensor field and is the building block of the interaction between the two metric fields. Finally, $e_n(\mathbb{X})$ denotes the $n$\textsuperscript{th} so-called elementary symmetric polynomial of the eigenvalues of $\mathbb{X}$, written most compactly as
\begin{equation}
  \begin{gathered}
  	e_0 = \mathbb{1},\quad e_1 = \tr{\mathbb{X}},\quad e_2 = \frac{1}{2} \left[ \tr{\mathbb{X}}^2 - \tr{\mathbb{X}^2}\right],\\
    e_3 = \frac{1}{6} \left[ \tr{\mathbb{X}}^3 - 3\, \tr{\mathbb{X}}\tr{\mathbb{X}^2}+2\, \tr{\mathbb{X}^3}\right],\quad
    e_4 = \mathrm{det}(\mathbb{X}).
  \end{gathered}
\end{equation}
The variation of this action yields two sets of Einstein equations,
\begin{subequations}\label{eq:Einstein}
  \begin{align} 
   &  G(g)_{\mu\nu} + m^2 \sin^2(\theta)\sum_{n=0}^3 \beta_n V^{(n)}(g)_{\mu\nu} = 8 \pi G_N T_{\mu\nu} \\
 &    G(f)_{\mu\nu} + m^2  \cos^2(\theta)  \sum_{n=1}^4 \sqrt{\frac{|\det g|}{|\det f|}} \beta_n V^{(n)}(f)_{\mu\nu} = 0, 
  \end{align}
\end{subequations}
where $  \sin^2(\theta) \equiv \frac{M_\text{eff}^2}{M_g^2} $, $  \cos^2(\theta) \equiv \frac{M_\text{eff}^2}{M_f^2} $ , and $8 \pi G_N = M_g^{-2}$ is the relation between Newton's constant and the (reduced) Planck mass for $g$. Moreover, the interaction or mass terms $V^{(n)}(g)$ follow from the variation of the polynomials $e_n$:
\begin{subequations}
  \begin{align}
  	{V^{(0)}(g)^\mu}_\nu =& {\delta^\mu}_\nu\\
    {V^{(1)}(g)^\mu}_\nu =& \tr{\mathbb{X}}{\delta^\mu}_\nu - {\mathbb{X}^\mu}_\nu ,\\
    {V^{(2)}(g)^\mu}_\nu =& {\left(\mathbb{X}^2\right)^\mu}_\nu - \tr{\mathbb{X}} {\mathbb{X}^\mu}_\nu + \frac{{\delta^\mu}_\nu}{2}\left[\tr{\mathbb{X}}^2 - \tr{\mathbb{X}^2}\right],\\
    {V^{(3)}(g)^\mu}_\nu =& -{\left(\mathbb{X}^3\right)^\mu}_\nu + \tr{\mathbb{X}} {\left(\mathbb{X}^2\right)^\mu}_\nu
			  - \frac{1}{2} \left[\tr{\mathbb{X}}^2 - \tr{\mathbb{X}^2}\right]{\mathbb{X}^\mu}_\nu +\nonumber \\
			  & + \frac{{\delta^\mu}_\nu}{6} \left[ \tr{\mathbb{X}}^3 - 3\, \tr{\mathbb{X}}\tr{\mathbb{X}^2}+2\, \tr{\mathbb{X}^3}\right].
  \end{align}
\end{subequations}
The corresponding expressions $V^{(1,2,3)}(f)$ are obtained from the $V^{(2,3,4)}(g)$ by dropping the parts proportional to ${\delta^\mu}_\nu$ and multiplying by $(-1)$, while for $n=4$, one obtains ${V^{(4)}(f) ^\mu}_\nu= {\delta^\mu}_\nu$.

Finally, energy-momentum conservation can be enforced by demanding the vanishing of the $g$-covariant derivative of the $g$-interaction term,
\begin{equation} \label{eq:Bianchi}
  \nabla(g)_\mu {T^\mu}_\nu\stackrel{!}{=} 0 \ \Rightarrow \ \nabla(g)_\mu {V^{(n)}(g)^\mu}_\nu = 0.
\end{equation}
Equivalently, we find for the $f$-metric $\nabla(f)_\mu \left(\sqrt{\frac{|\det g|}{|\det f|}} {V^{(n)}(f)^\mu}_\nu\right) = 0$. These additional equations are known as Bianchi constraints, which are in general \emph{not} independent. 
We now show the phenomenological consequences of the bigravity extension for the non-relativistic gravitational potential and gravitational lensing effects. 

\subsection{Gravitational Potential in Bigravity}
The effective Newtonian potential felt by a non-relativistic observer in a static and spherically symmetric potential of a point source of mass $M$ has been calculated in~\citep{Babichev:2015xha, Platscher:2016adw} and reads
\begin{equation}
\label{Eq:FullPotential}
\phi(r) =
\begin{cases}
 - \frac{M G_N}{r}   & r \ll r_V,
 \\
- \frac{M G_N}{r} \left[  \alpha(\theta) + \beta(\theta) e^{- m_gR}\right]  & r \gg r_V,
\end{cases}
\end{equation}
with $\alpha(\theta) := \cos^2(\theta)\left(1+\frac{2}{3}\sin^2(\theta)\right)$, $\beta(\theta) :=  \frac{2}{3} \sin^2(\theta) \left(1 + 2 \sin^2(\theta)\right)$, and $m_g^2  = m^2 (\beta_1 + 2 \beta_2 + \beta_3)$. In the above expression, $r_V$ is the scale below which non-linearities become important and conspire to reproduce the GR result, as conjectured by Vainshtein in 1972~\cite{Vainshtein:1972sx}. Below, we will assume that, as we lower $r$ below $r_V$, the two regimes are smoothly interpolated. As a working hypothesis, we propose an effective mixing angle to account for this phenomenon, which is defined as follows:
\begin{equation}\label{eq:Vainshtein}
	\theta_\text{eff} \equiv \frac{\theta}{2} \left[1 + \tanh \left(\frac{r - r_V}{\Delta r}\right) \right],
\end{equation}
where $\Delta r $ is treated as a free parameter in our analysis. This parameter encodes the effect of the extended matter distribution on the Vainshtein mechanism, which are difficult to calculate from first principles. Thereby, we find that for $r \gg r_V$, $\theta_\text{eff} = \theta$ and for $r \ll r_V$, we have that $\theta_\text{eff} = 0$, i.e.~GR is recovered. Finally, we use the following expression for the Vainshtein radius, motivated by the study of spherically symmetric solutions, cf.~Refs.~\cite{Babichev:2009us,Babichev:2009ee,Babichev:2009jt,Babichev:2010jd,Babichev:2015xha,Platscher:2016adw}: 
\begin{equation}\label{eq:rVdef}
	r_V = \sqrt[3]{\frac{r_S}{m_g^2}}\,,
\end{equation}
with the Schwarzschild radius $r_S = 2G_N M$. For scales $r \ll r_V$ one finds that non-linearities become important and conspire to restore GR.\footnote{An interesting observation is that this exact screening could be broken and thereby be testable~\cite{Salzano:2017qac}.} The regime $r \gg r_V$ turns out to be well approximated by a linearised treatment.  We will discuss the finite matter extension effects in Sec.~\ref{sec:rotCurves}.

\subsection{Bending of light in Bigravity \label{sec:lensing}}

As discussed above the metric and thus the lensing potential in bigravity are exactly the same as in GR inside the Vainshtein sphere. However, outside we find the deviation by considering the metric around a spherical body in a linear approximation~\cite{Babichev:2015xha}, which can be written as 
\begin{equation}
\begin{gathered}
ds^2 = -dt^2 e^\nu + e^\lambda d r^2 + r^2 \mathrm{d}\Omega^2\,, \quad  \text{with}\\
 \nu(r) = -\frac{r_S}{r} \left( \alpha(\theta) + \beta(\theta) e^{-m_g r}  \right)
\quad\text{and}\quad  \lambda(r) = \frac{r_S}{r} \left( \alpha(\theta) + \frac{\beta(\theta) (1+ m_g r)}{2} e^{-m_g r}  \right) \,. 
\end{gathered}
\end{equation}
The null-geodesic equation for an impact parameter $b$ is readily derived from this and reads
\begin{equation}\label{eq:light_geodesic0}
	\left(\frac{1}{r^2} \frac{\mathrm{d}r}{\mathrm{d}\phi}\right)^2 = \frac{ e^{-(\lambda+\nu)}}{b^2} -  \frac{e^{-\lambda}}{r^2} \,.
\end{equation}
To solve this equation, we introduce the variable $u \equiv \frac{R}{r}$, where $R$ is the radius of closest approach, obtained by setting  $\left.\frac{\mathrm{d}r}{\mathrm{d}\phi}\right|_{r=R} = 0$, and which is related to $b$ via
\begin{equation}
	\frac{1}{b^2} = \frac{e^{\nu}}{R^2}\, .
\end{equation}
Finally, expanding the exponentials in $\frac{r_S}{R}$, we arrive at the geodesic equation,
\begin{equation}\label{eq:light_geodesic}
	\left(\frac{\mathrm{d}u}{\mathrm{d} \phi}\right)^2 = \big[1 - \lambda (r=R/u) \big] (1 - u^2) + \nu(R) - \nu(r=R/u)\,.
\end{equation}
Furthermore, the final expression for the light deflection angle is given by
\begin{equation} \label{eq:deltaPhi_1}
	\Delta \phi = 2\, \int \mathrm{d} \phi = 2 \, \int_0^1 \mathrm{d} u\, \frac{\mathrm{d} \phi}{\mathrm{d} u}\,,
\end{equation}
which can be solved in principle via Eq.~\eqref{eq:light_geodesic}. In GR one makes use of the fact that $e^\nu \, e^\lambda =1$, which is no longer true in bigravity; in fact, $e^\nu \, e^\lambda \sim e^{-m_g r}$, i.e.~it is the piece that carries the information about the massive spin-2 mode. In order to arrive at a final result, we plug Eq.~\eqref{eq:light_geodesic0} into~\eqref{eq:deltaPhi_1} before expanding in powers of $\frac{r_S}{R}$. 

\begin{figure}
	\centering
	\includegraphics[width=\textwidth]{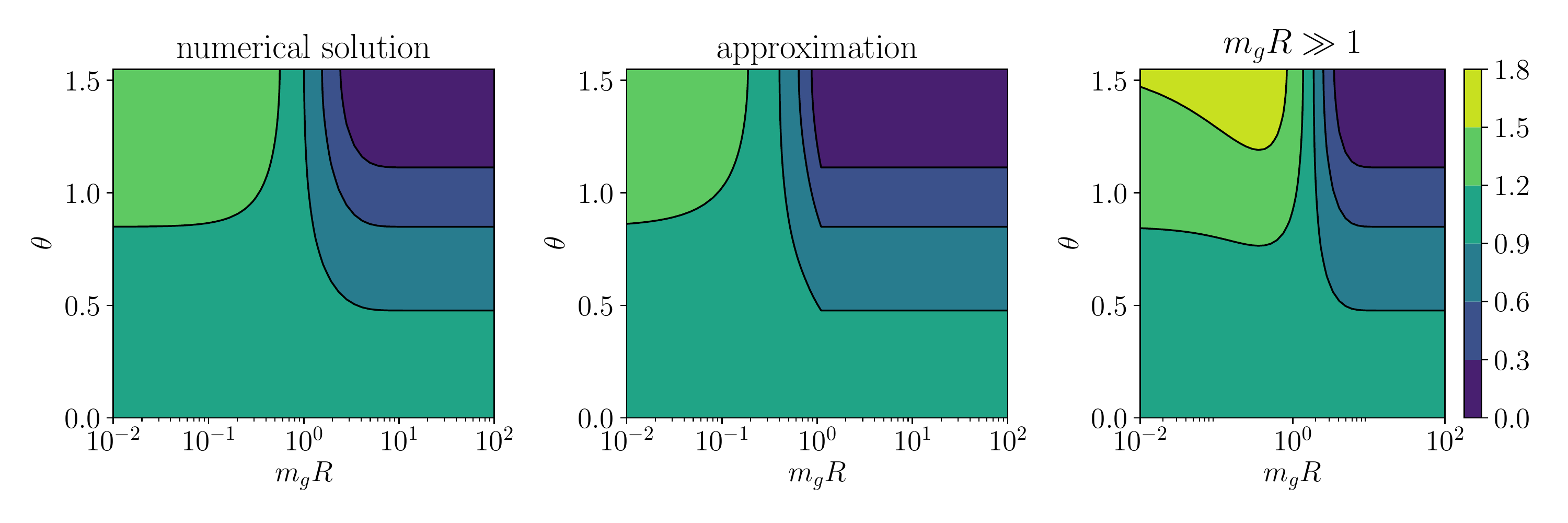}
	\caption{\label{fig:lensing_test}The ratio of the deflection angle of light in bigravity and GR with the two different approximate solutions. All expressions agree when $m_g R \gg1$, however only the careful approximation~\eqref{eq:deltaPhi_2} gives a good estimate of $\Delta \phi$ when $m_gR < 1$.}
\end{figure}

Unfortunately, we cannot give a closed form expression for $\Delta \phi$ obtained from the integration of Eq.~\eqref{eq:deltaPhi_1} due to the $u$-dependent  exponentials in the integrand. However, one can find a satisfactory approximation by observing that the exponentials of the form $e^{-m_g\, R /u}$ are only relevant in the case where $m_gR <1$. In this case, they are close to unity when the integration variable $u\to 1$, and approach zero as $u < m_gR$. We capture this behaviour by replacing these exponentials as $e^{-m_g\, R /u} \mapsto e^{-m_g\, R}$, and shifting the lower integration boundary from $u=0$ to $u=m_gR$. The resulting expression reads
\begin{equation} \label{eq:deltaPhi_2}
	\Delta \phi \simeq 2 \frac{r_S}{R} \left[ \alpha(\theta) + \frac{1}{4} \beta(\theta) e^{-m_gR} \left( \left(3 + m_gR\right) \sqrt{\frac{1+ m_gR}{1- m_gR}} + m_gR\, \arccos \left(m_gR\right)\right)   \right]\,.
\end{equation}
Note that this expression develops an imaginary part when $m_g R >1$; however, one can safely disregard the part proportional to $\beta(\theta)$ in this regime, e.g.~numerically via a Heaviside step-function. Another approximation one can make is $m_g R \gg 1$, keeping only the exponential in Eq.~\eqref{eq:deltaPhi_2}, which yields
\begin{equation}\label{eq:deltaPhi_3}
	\Delta \phi \simeq 2 \frac{r_S}{R} \left[ \alpha(\theta) + \frac{3}{4} \beta(\theta) e^{-m_gR} \right]\,.
\end{equation}
We compare the different results for $\Delta \phi$ in Fig.~\ref{fig:lensing_test} and conclude that Eq.~\eqref{eq:deltaPhi_2} gives a fair approximation to the numerical solution, while Eq.~\eqref{eq:deltaPhi_3} does not reproduce the numerical result for $m_gR <1$	--~as was to be expected.

\section{Constraints from Galaxy Clusters}
\label{sec:clusters}

Our goal is to break the degeneracy of the effect of bigravity and the potential effect of a DM component. To this end we study physical systems of different size and mass. We will begin by investigating the effect of bigravity in the largest gravitationally bound systems in the Universe, namely galaxy clusters. As we will illustrate now for bigravity, the ratio of kinematic mass and lensing mass estimates differs from unity, which is never the case in GR. Thus we are able to set limits on the bigravity model, without any assumptions on the mass-to-light ratio of the object. 

\subsection{Gravitational Lensing}

Depending on the graviton mixing angle $\theta$, the deflection angle of light changes according to
\begin{equation}
  \Delta\phi \to \Delta\phi\,f_\mathrm{GL}(\theta)
\label{eq:1}
\end{equation}
where $  f_\mathrm{GL}(\theta)$ is found by performing the integral in Eq.~\eqref{eq:deltaPhi_1} and is well approximated by
\begin{equation}
 \alpha(\theta) + \frac{1}{4} \beta(\theta) e^{-m_gR} \left( \left(3 + m_gR\right) \sqrt{\frac{1+ m_gR}{1- m_gR}} + m_gR\, \arccos \left(m_gR\right)\right) \,,\quad
  0\le\theta\le\frac{\pi}{2}\,,
\label{eq:2}
\end{equation}
as we have seen in the previous section. Since this function does not depend on the radius, derivatives of the deflection angle (such as lensing shear and convergence) will change by the same factor $f_\mathrm{GL}$. Mass estimates from gravitational lensing (taken outside the Vainshtein radius) will hence change as
\begin{equation}
  M_\mathrm{lens} \to \frac{M_\mathrm{lens}}{f_\mathrm{GL}}\;.
\label{eq:2a}
\end{equation} 
Thus, observing the strength of the lensing effect, we can reconstruct the lensing mass responsible for the light bending $M_\text{lens}$, which depends on the bigravity parameters $m_g$ and $\theta$. We will now turn to a method to reconstruct the gravitational potential in a cluster, by X-ray observations to have an independent observable of the induced metric.

\subsection{X-ray Emission of Galaxy Clusters}

As discussed above the gravitational potential is replaced by
\begin{equation}
  \phi \to \phi\,f_\phi(\theta, r)
\label{eq:3}
\end{equation}
with
\begin{eqnarray}
  f_\phi(\theta, r) &:=& \cos^2\theta\left(1+\frac{2}{3}\sin^2\theta\right)+
  \frac{2}{3} \sin^2 \theta \left(1+2\sin^2 \theta\right)\,e^{-m_g r}
  \nonumber\\ &\,=&
    \alpha(\theta)+\beta(\theta)\,e^{-\lambda r}\;.
\label{eq:4}
\end{eqnarray}
In hydrostatic equilibrium,
\begin{equation}
  \frac{\vec\nabla P}{\rho} = -\vec\nabla\phi\;.
\label{eq:5}
\end{equation}
According to (\ref{eq:3}),
\begin{equation}
  -\vec\nabla\phi \to -\vec\nabla\phi\,f_\phi-\phi\,\vec\nabla f_\phi\;.
\label{eq:6}
\end{equation}
In spherical symmetry and ordinary gravity,
\begin{equation}
  \phi = -\frac{GM(r)}{r}\;;\quad
  -\vec\nabla\phi = -\frac{GM(r)}{r^2}\;,
\label{eq:7}
\end{equation} 
and thus from (\ref{eq:6})
\begin{equation}
  \frac{GM(r)}{r^2} \to 
  \frac{GM(r)}{r^2}\,\left(f_\phi-r\D_rf_\phi\right) =
 \frac{GM(r)}{r^2}\,\left(\alpha+\beta(1+\lambda r)e^{-\lambda r}\right)\;.
\label{eq:8}
\end{equation}
For an ideal gas,
\begin{equation}
  \vec\nabla P = P\left(\vec\nabla\ln\rho+\vec\nabla\ln T\right)\;,
\label{eq:9}
\end{equation}
so that hydrostatic equilibrium (\ref{eq:5}) requires
\begin{equation}
  -\frac{k_\mathrm{B}T}{\bar m\,r}
  \left(\frac{\D\ln\rho}{\D\ln r}+\frac{\D\ln T}{\D\ln r}\right) =
  \frac{GM(r)}{r^2}\,\left(\alpha+\beta(1+\lambda r)e^{-\lambda r}\right)\;.
\label{eq:10}
\end{equation}
Since the left-hand side of this equation is measured and thus remains unchanged, Eq.~\eqref{eq:10} implies that the mass inferred from X-ray data will change according to
\begin{equation}
  M_\mathrm{kin}(r) \to
\frac{M_\mathrm{kin}(r)}{\alpha+\beta(1+\lambda r)e^{-\lambda r}}\;.
\label{eq:11}
\end{equation}
Therefore, we can determine the effective underlying parameter $M_\text{kin}$ which sources the gravitational potential. The ratio of cluster masses inferred from X-rays and gravitational lensing is therefore expected to change according to
\begin{equation}
  \left\langle\frac{M_\mathrm{kin}(r)}{M_\mathrm{lens}(r)}\right\rangle \to
  \left\langle\frac{M_\mathrm{kin}(r)}{M_\mathrm{lens}(r)}\right\rangle\,
    \frac{f_\mathrm{GL}}{\alpha+\beta(1+\lambda r)e^{-\lambda r}} \equiv R_M \left\langle\frac{M_\mathrm{kin}(r)}{M_\mathrm{lens}(r)}\right\rangle\;.
\label{eq:12}
\end{equation}
This ratio $R_M$ contains non-trivial  information about both principal functions  of the induced metric  $\lambda$ and $\nu$. It will be useful to search for deviations from the GR result, which is unavoidably $R_M=1$. 

Notice that in order to arrive at Eq.~\eqref{eq:12}, we have to assume that light propagates outside the Vainshtein sphere, which turns out to be a sensible approximation for the physical systems under consideration. Otherwise a more involved analysis of the gravitational potential inside cluster is needed which is beyond the scope of this work, cf.~Appendix~\ref{app:potentials}.

\subsection{Identifying Clusters against the CMB}

The CMB temperature fluctuations normalise the small-scale density fluctuations via the shape of the power spectrum. Thus, the expected number of galaxy clusters can be predicted from it. Via the thermal  Sunyaev -- Zel'dovich (SZ) effect, clusters can directly be seen against the CMB. The inverse Compton scattering off high energy electrons in clusters causes distortions of the CMB spectrum. However, the observed number of clusters is substantially lower than that expected. Since the thermal SZ effect needs hot gas, one solution could be that clusters of a given mass are cooler than expected. Tentatively reconciling this conclusion with hydrostatic equilibrium, there must be non-thermal pressure support, or a deviation from the relation $ M_{\rm kin}/M_{\rm lens} =1$. This is quantified by the hydrostatic bias factor introduced above, which needs to be smaller than unity, $R_M = 0.66 \ldots 0.99 $ (depending on the prior chosen), cf.~Ref.~\cite{Ade:2015fva}.  

\subsection{Bigravity effect on observations of lensing in clusters}

\begin{figure}[t]
	\centering
	\includegraphics[width=.75\textwidth]{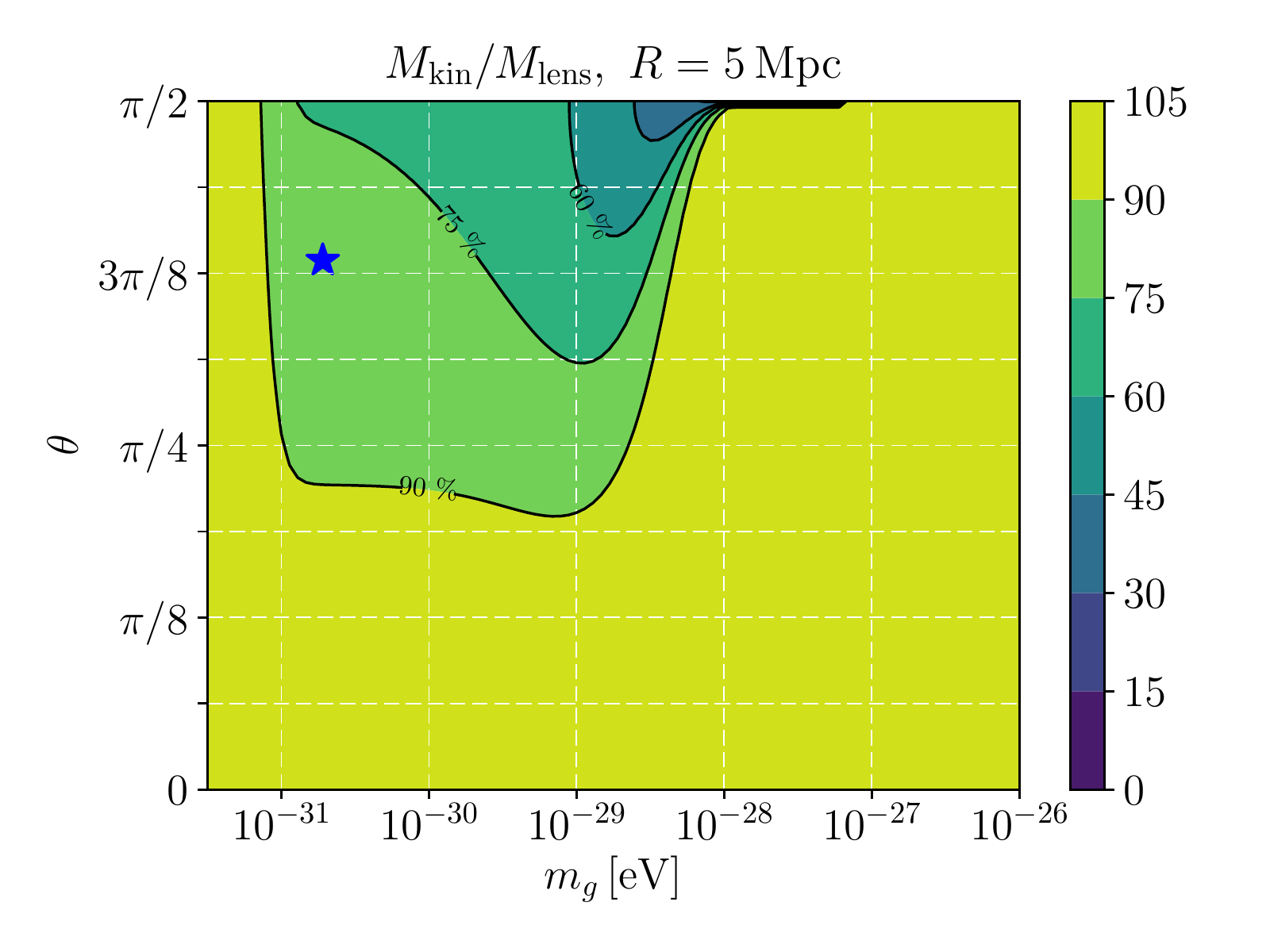}
	\caption{\label{fig:Contours}Contour plot of the ratio $\left\langle\frac{M_\mathrm{kin}}{M_\mathrm{lens}}\right\rangle$ relative to the GR value for a cluster size of $5\Mpc$ in percent. The blue star indicates the best fit point, favoured by data from the cluster J1206.2-0847 studied below.}
\end{figure}

Combining the results of the previous subsections, we can explore the effects of the modified potentials by producing Fig.~\ref{fig:Contours}, where the ratio $R_M$ is shown for a cluster of radius $R_c = 5\Mpc$. We quote the aforementioned value for the bias parameter $R_M = 0.66 \ldots 0.99 $, suggested by CMB observations. 

We see that in a region where the mass of the massive spin-2 mode lies roughly between $10^{-31}\eV$ and  $10^{-28}\eV$, a reduction of the ratio $R_M$ is indeed feasible for large ($\theta \gtrsim \pi/4$) mixing angles. Moreover, in the limit of pure massive gravity ($\theta = \pi/2$), there is a mass window $10^{-28}\eV \lesssim m_g \lesssim 10^{-27} \eV$, where the ratio nearly vanishes. Therefore, it remains an intriguing question whether the tension with the GR prediction of $R_M$ is indeed due to deviations from hydrostatic equilibrium or a hint for new physics in the gravitational sector. 

Finally, we remark that for masses well below $10^{-31}\eV$, the Vainshtein mechanism sets in, i.e.~$\theta_\text{eff}=0$, while for masses $m_g > 10^{-27}\eV$, we find that the Compton wavelength of the graviton, $\lambda_c = m_g^{-1} \ll R_c$ and the exponential $e^{m_g R_c} \to 0$. Consequently, the ratio $R_M = 1$ in both regimes.

\subsection{The MACS J1206.2-0847 Cluster}\label{sec:MACSlensing}

We now use data of the MAssive Cluster Survey (MACS) galaxy cluster J1206.2-0847, located at a redshift $z = 0.44$~\cite{Ebeling:2000nh,Ebeling:2009sn,Postman:2011hg,Annunziatella:2014vsa},  to explicitly demonstrate the effect of the deviation of the kinetic and lensing masses. In Ref.~\cite{Biviano:2013eia} two different techniques were used to obtain information about the kinematic mass $M_\text{kin}$, i.e.~the mass inducing the gravitational potential. Combining both, a reconstruction of the mass was possible within the viral radius as well as beyond it. The crucial part for our analysis is the comparison between the reconstructed $M_\text{kin}$ and the lensing mass $M_\text{lens}$ obtained for the same cluster in \cite{Annunziatella:2014vsa}. As discussed above,  outside the Vainshtein radius the ratio $R_M$ of these masses will deviate from one.  

\begin{figure}[t]
	\centering
	\subfloat[exclusion region and best-fit point]{
		\includegraphics[width=.47\textwidth]{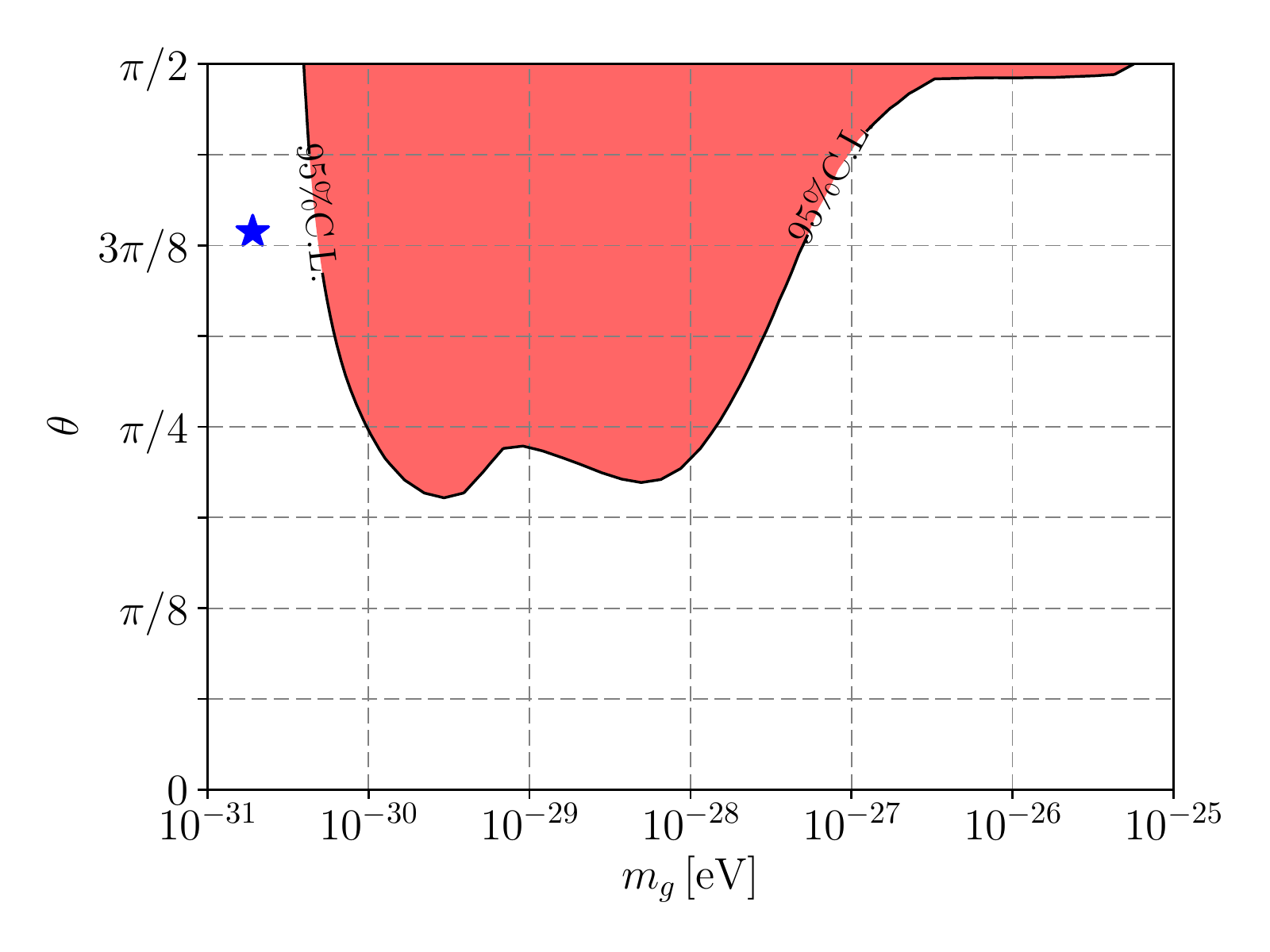}\label{fig:ClusterExample} }
	\subfloat[$R_M$ as function of the radius]	{
		\includegraphics[width=.47\textwidth]{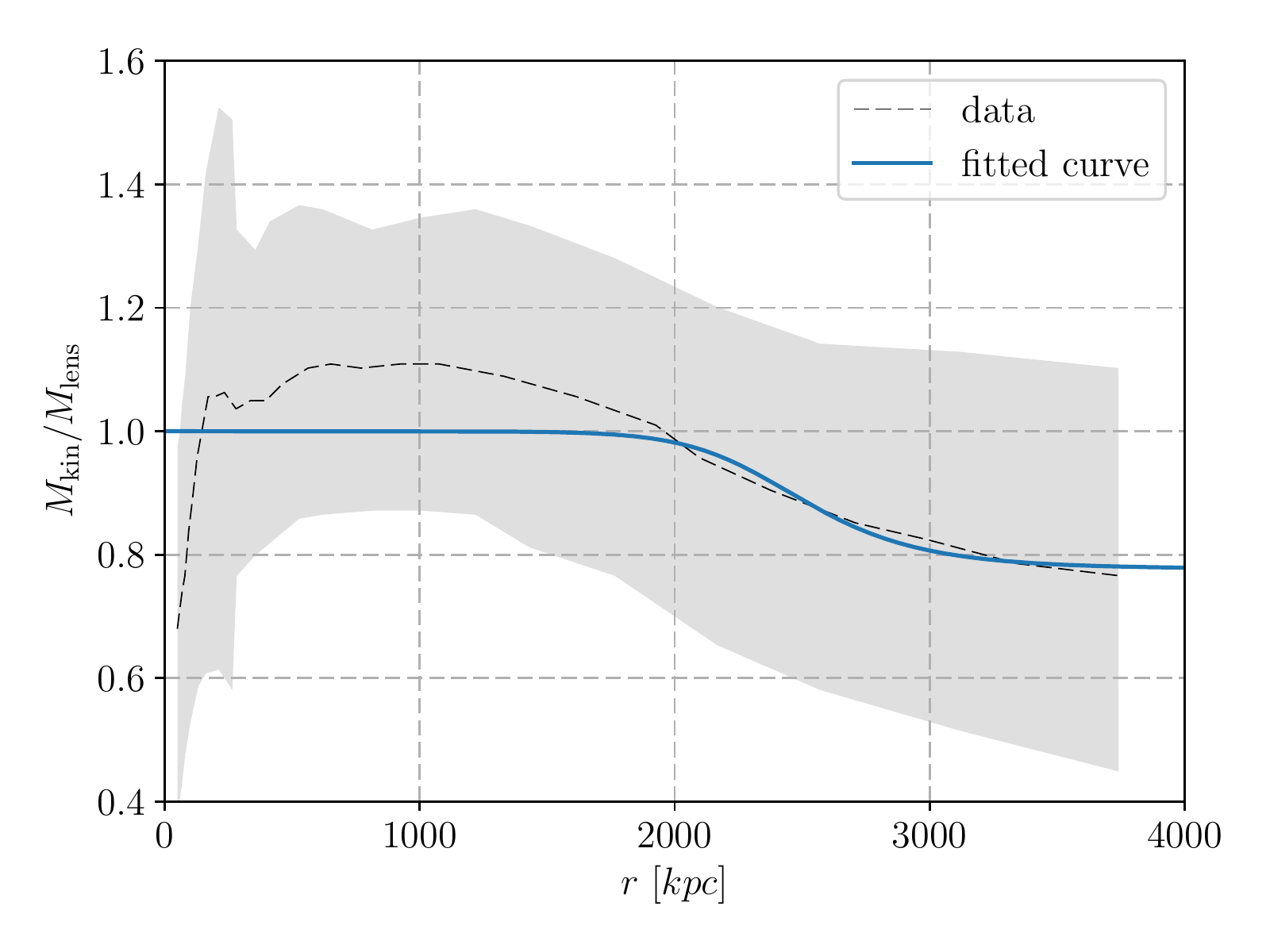}\label{fig:MACSfit} }
	\caption{\emph{Left:} $\chi^2$ parameter study in bigravity for the ratio $R_M$ of the MACS J1206.2-0847 cluster assuming the hypothesis of graviton mixing with a massive spin-2 state. The $\chi^2$ value is converted into an exclusion region at the 95\% C.L.~and a best-fitting point (blue star). \emph{Right:} 
	Ratio of kinetic to lensing mass $R_M$ of the  MACS J1206.2-0847 cluster as a function of radial distance, given the best fit parameters $m_g  = 2  \cdot 10^{-31}\eV,\ \theta =1.2,\ \Delta r = 0.23\, r_V,\ \Delta {\chi^2} \equiv (\chi^2 - \chi^2_0)/\chi^2_0 = -0.35$.}
\end{figure}

In order to draw quantitative conclusions, we compare the observations of $R_M$ and the model prediction of bigravity as a function of the two parameters graviton mass, $m_g$, and the graviton mixing angle, $\theta$.  We perform a $\chi^2$ analysis and show the results in Fig.~\ref{fig:ClusterExample}. We find that a large region of the mass window of interest is already ruled out at the $95\%$ confidence level (C.L.) when the mixing angle is large enough. This is consistent with previous studies of lensing constraints in pure massive gravity \cite{Choudhury:2002pu}. An intriguing observation is that for $m_g \gtrsim 10^{-31}\eV$  and large mixing angles the $\chi^2$ of the fit is reduced significantly relative to the null hypothesis $R_M =1$. For this particular choice of parameters, which yield the best fit as indicated by a blue star in Figs.~\ref{fig:Contours} and~\ref{fig:ClusterExample}, we present the fit to the data in Fig.~\ref{fig:MACSfit}. 

\subsection{The Bullet Cluster}

The main idea behind this analysis is the difference in the gravitational effect of compact structures and extended gravity sources in theories with a Vainshtein mechanism at work. 
Let us first define the radius $r_M$ within which the major matter fraction is concentrated. Furthermore,  under a certain graviton mass hypothesis we find the would-be Vainshtein radius given the enclosed mass $r_V(M_\text{in}, m_g)$. We distinguish two cases,   1) $r_M < r_V(M_\text{in}, m_g)$ and 2) $r_V(M_\text{in}, m_g) < r_M $. In case 1) the Vainshtein radius is equivalent to the one found in the Schwarzschild spacetime, while in case 2) the matter backreaction effect stops the longitudinal mode from propagation within $r_M$, as discussed in Ref.~\cite{Babichev:2013usa}.  Thus we find that the effective region where GR is restored is within the $r_V^\text{induced} = \max ( r_V(M_\text{in}, m_g) , r_M)$. 

The two galactic core components of the bullet cluster~\cite{Tucker:1998tp}, correspond to the case 1). On the other hand, the plasma component, which is located at the center of the cluster collision and is displaced due to ram pressure, is not sufficiently compact and corresponds to case 2). As mentioned above, the matter coupling in the plasma clouds leads to the non-propagation of the longitudinal polarisation of the graviton. Thus, within the plasma cloud, GR is restored in the same way, as in the Vainshtein sphere around the more compact stars. For the our analysis we assume that within the region approximately homogeneously filled with gas, the gravitational law of massless gravity holds. 

\begin{figure}[t]
	\centering
	\subfloat[GR limit $\theta  = 0 $]	{
		\includegraphics[width=.45\textwidth]{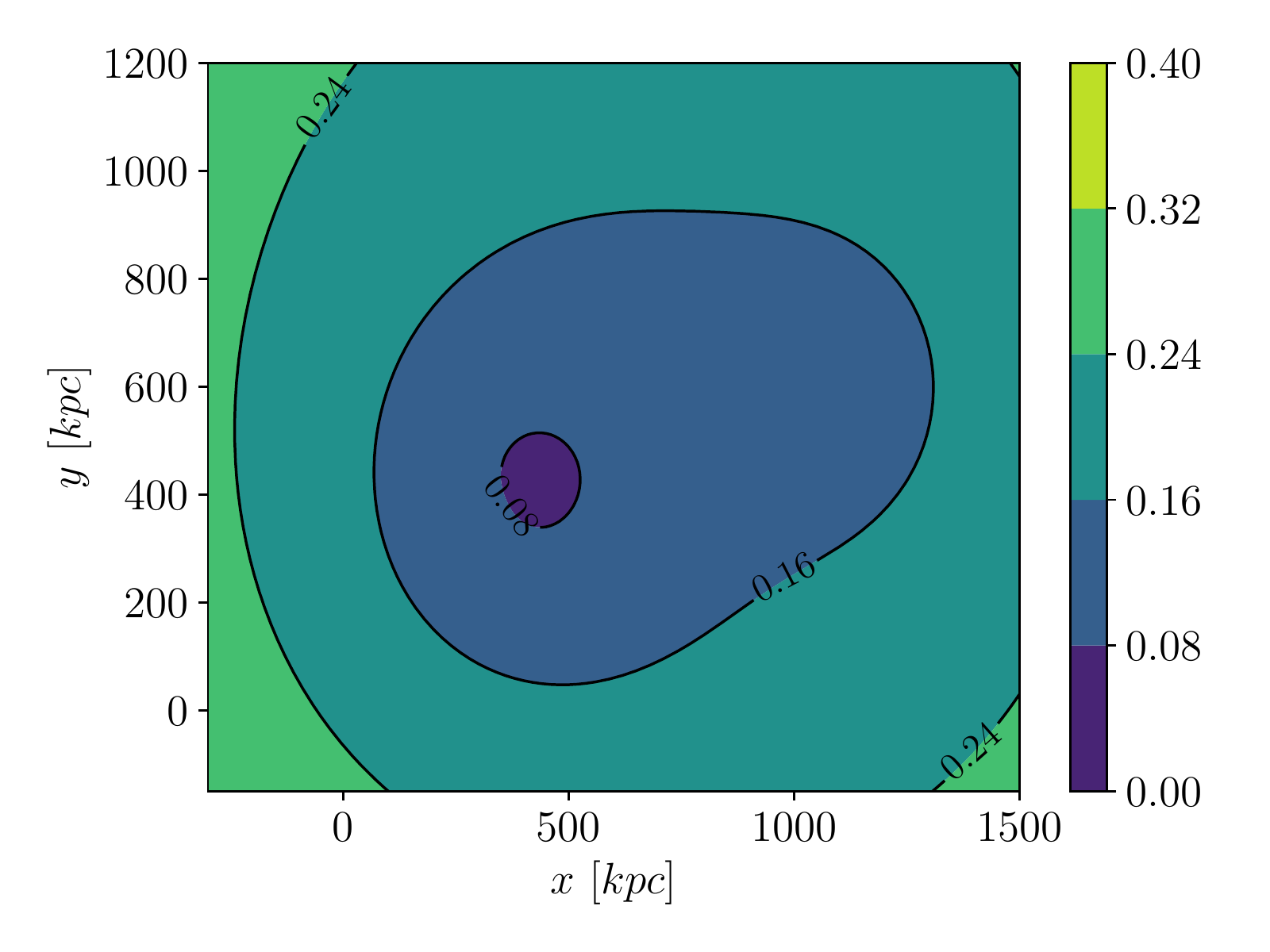}}
	\subfloat[$(m_g, \theta)=(10^{-30}\eV, \pi/4)$]{
		\includegraphics[width=.45\textwidth]{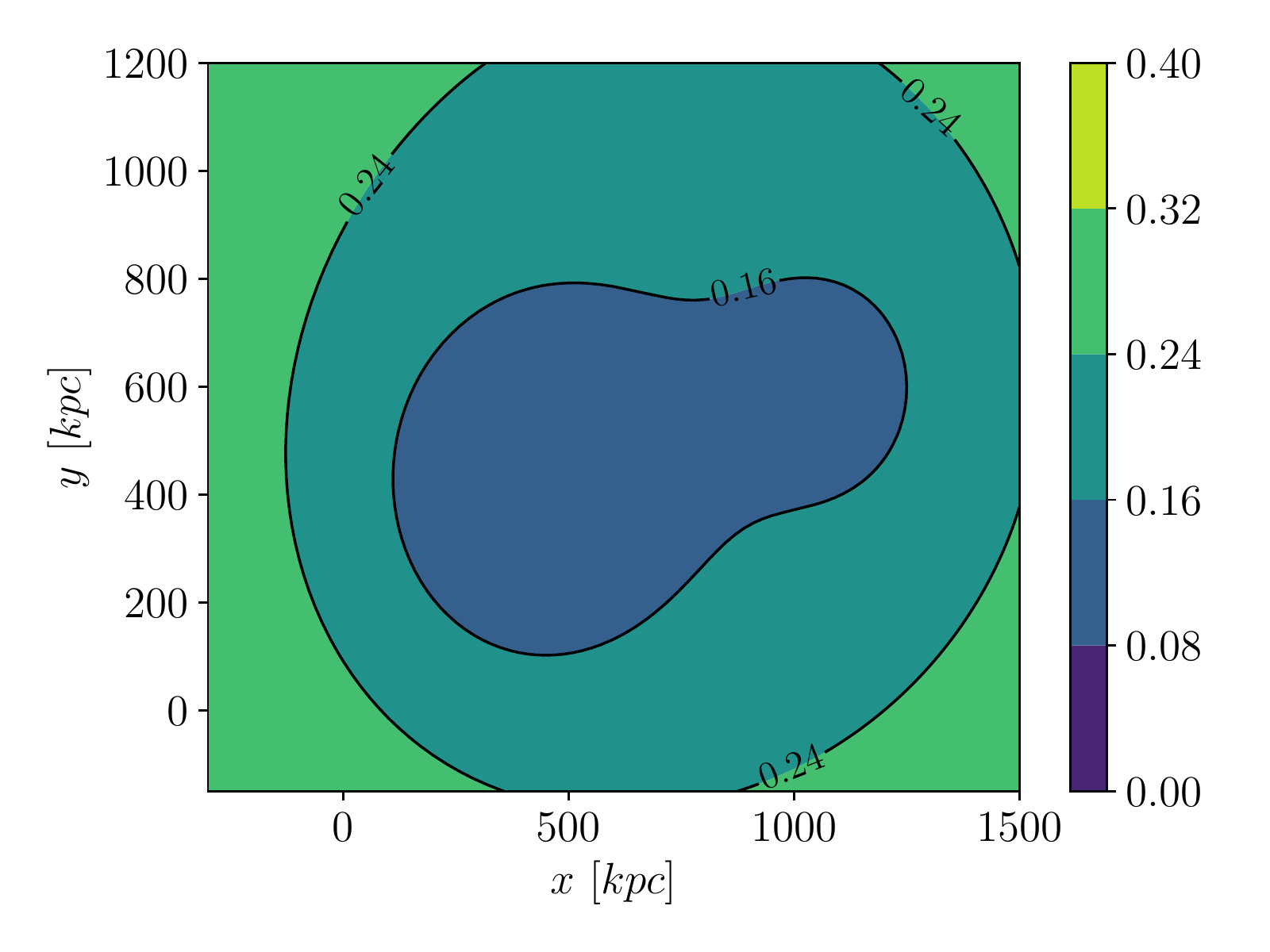}}\\
			\subfloat[$(m_g, \theta)=(10^{-30}\eV, \pi/3)$]{
		\includegraphics[width=.45\textwidth]{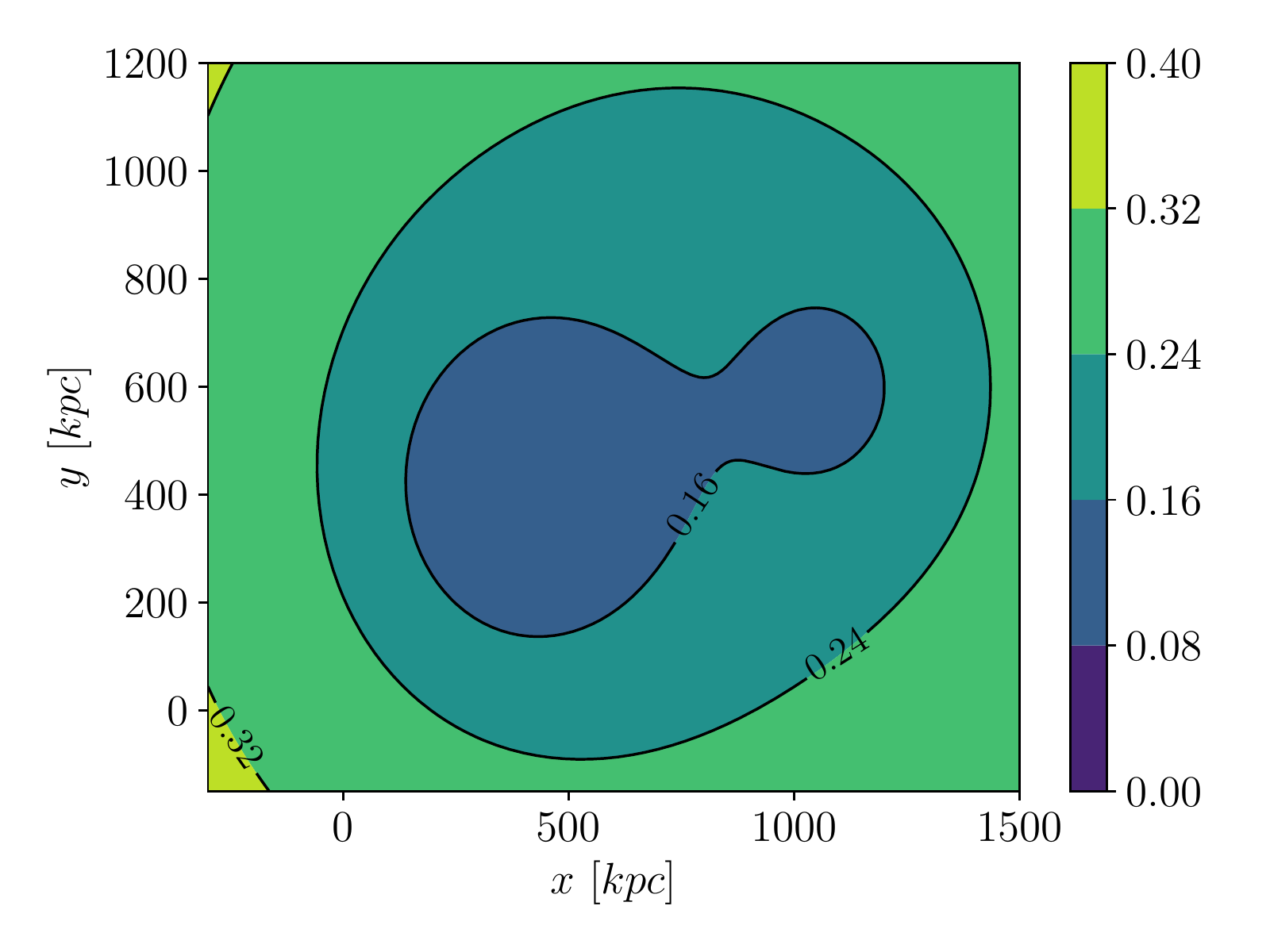}}
	\subfloat[$(m_g, \theta)=(10^{-30}\eV, \pi/2)$]{
		\includegraphics[width=.45\textwidth]{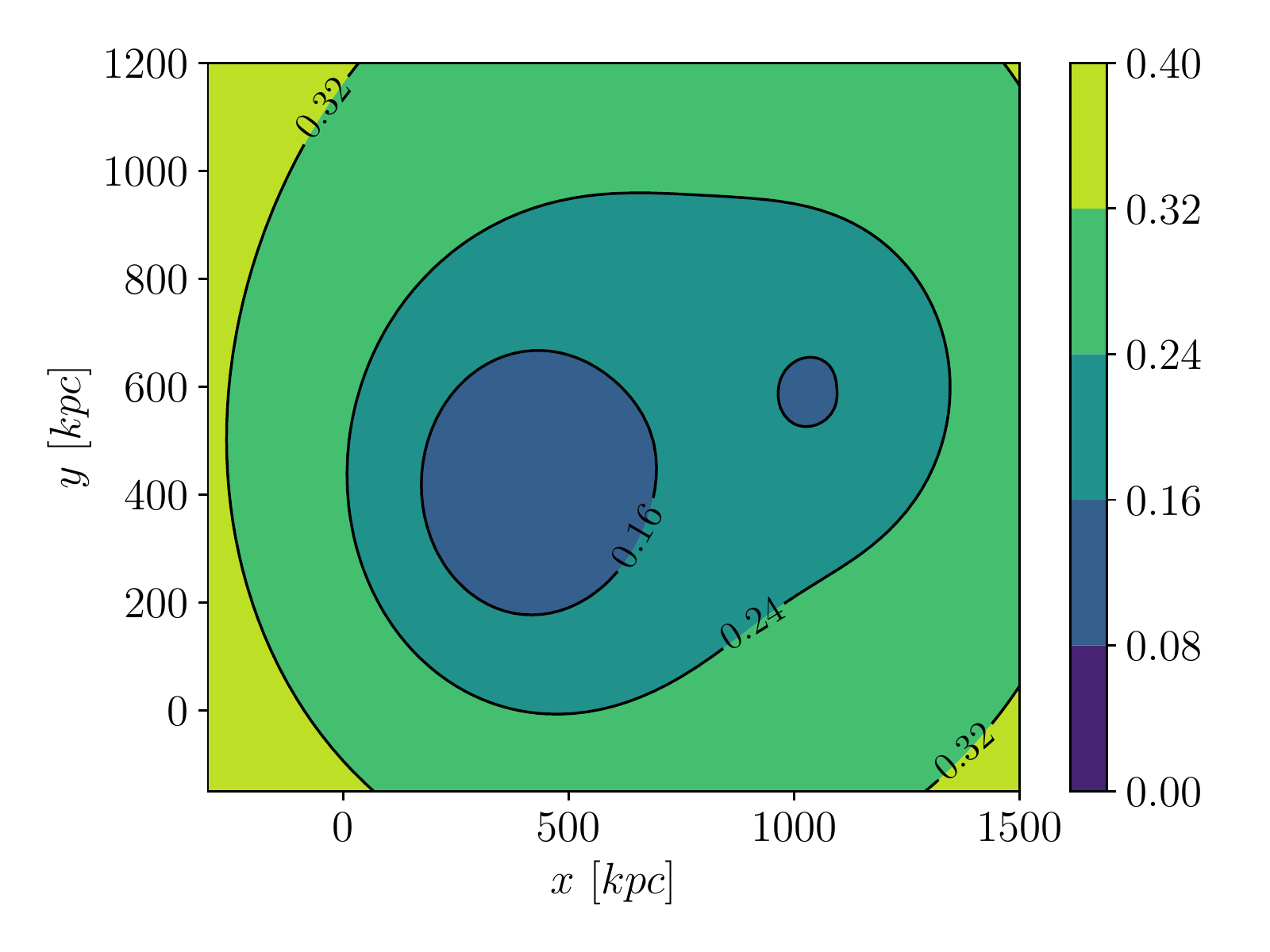}}
	\caption{ \label{fig:bullet} The parameter $\kappa$, which is the lens convergence and is a good approximation to the surface mass density of the mass distribution in the Bullet Cluster.  The four panels show $\kappa$ reconstructed under different hypotheses of the lensing potential (a) GR and (b,c,d) bigravity. }
\end{figure}

In Fig.~\ref{fig:bullet} we show how the mass of the Bullet cluster, reconstructed from weak lensing~\cite{Markevitch:2003at,Clowe:2003tk} changes when the effective mixing angle in bigravity is varied. In particular assuming only the stellar and plasma masses as sources of gravity, the surface mass density is computed in pure GR (Fig.~\ref{fig:bullet} a)), which is well approximated by the lens convergence $\kappa$. Then in  (Fig.~\ref{fig:bullet} b),c),d)) the massive mode mixing in bigravity is included and the lens convergence is modified according to the prescription described in section \ref{sec:lensing}. The enhanced lens convergence outside the Vainshtein sphere of the compact stellar source leads to a stronger apparent lensing potential. This would lead to a larger perceived mass surface density around the stars than in GR, but does not affect the lensing of the plasma cloud.   
In the regime with significant graviton mixing, it becomes obvious that indeed the displacement of the largest mass fraction, with respect to the plasma, is due to the fact that galaxies are collisionless. However, the modification of the lensing potential in bigravity lets their lensing mass appear higher at large scales than expected in GR. The qualitative picture seems to match the observations presented in Ref.~\cite{Clowe:2006eq}.

The above observation can be formulated in the following way. The Vainshtein radius of a baryon dominated system scales as $ r_V = (2 G_N \lambda_g^2 M_b)^{1/3}$, where $M_b$ is the baryonic mass. It can thus be written in terms of the baryonic density as $ r_V = (8 \pi/ 3 \, G_N \lambda_g^2 \rho_b)^{1/3} r_b$. The relevant ratio is $R_V = r_V/r_b$. This ratio has to be larger than one  $(R_V > 1$) in order for the Vainshtein radius to be outside the region where non-linear matter back-reaction restores GR. Thus, given a fixed graviton mass, there is a minimal baryonic matter density needed in order to induce observable deviations from GR. For example for a graviton mass of $10^{-30}$ eV, systems with lower densities than $2.5 \cdot 10^5\, M_\odot/\text{kpc}^3$ would be completely GR dominated. Therefore, we find that in bigravity the existence of systems which lack the DM phenomenon could occur naturally and is an interesting tool to constrain the graviton parameter space. We will discuss the impact of a recently observed, baryon-dominated galaxy on the graviton parameter space in the next section. 

\section{Galaxy Rotation Curves} \label{sec:rotCurves}
The previous section showed that, due to the Vainshtein mechanism, gravitational interactions are modified in the framework of bigravity on extra-galactic scales. Given that the Vainshtein radius itself is a function of the Schwarzschild radius, and hence of the mass of the system under consideration, one is led to ask if galactic dynamics are influenced, too. In the following we try to answer this question by means of galactic rotation curves; however, we emphasise that conclusive statements could only be drawn from a survey of many galaxies, while in this work we study only a few examples. We proceed as follows: first, we find a model for a galaxy that gives a good fit to the data with GR interactions and a DM halo, which we take to have a profile as proposed by Navarro, Frenk and White (NFW)~\cite{Navarro:1996gj}. We then turn on bigravity, i.e.~choose a non-zero mixing angle and graviton mass and repeat the analysis. We then quantify how much particle DM is needed with respect to GR. Interestingly, we find that the best-fitting point is one with reduced DM content, a non-zero mixing angle, and a graviton mass within one order of magnitude of the best-fitting point from the previous section.

Let us now go through the details. We demonstrate our analysis using  the ESO138-G014 galaxy, which was observed at a distance of 18.57~Mpc~\cite{OBrien:2010dps,Aaronson1982} and has a well measured velocity profile, see also Ref.~\cite{Hashim:2014mka}. Moreover, we perform a fit to a set of low surface brightness galaxies from Ref.~\cite{KuziodeNaray:2007qi} to derive bounds on the bigravity parameter space. 

\begin{figure}[t]
	\centering
	\subfloat[rotation curve parameter scan with 95$\%$ exclusion limits from the $\chi^2$ distribution. Best-fit point (red star) and the least DM point (yellow star).]{
	\includegraphics[width=.45\textwidth]{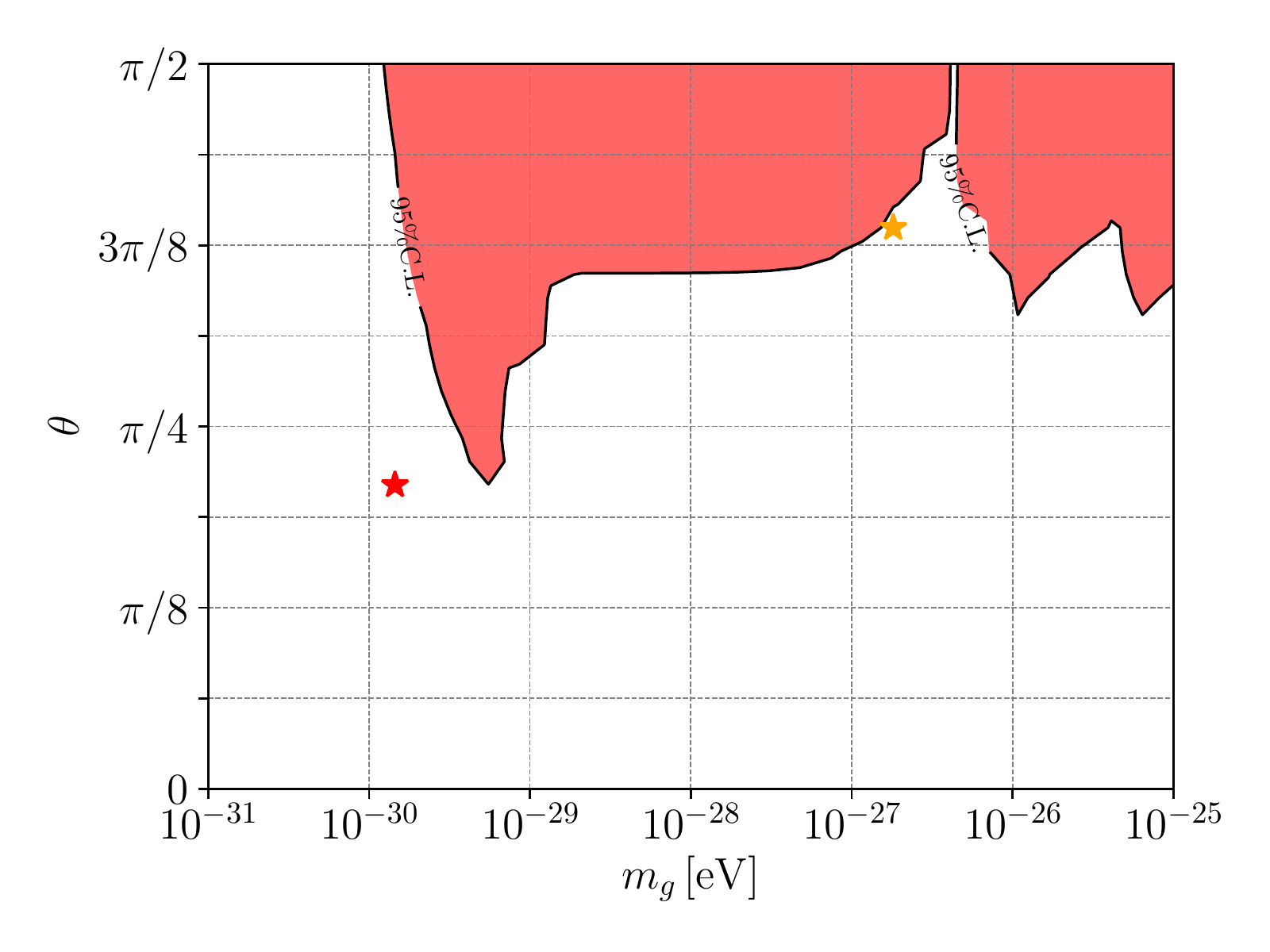}\label{fig:MassratioChisquared}}
	\subfloat[	mass ratio of DM to baryonic matter consistent with the rotation curves, within the central galactic region.]{
	\includegraphics[width=.47\textwidth]{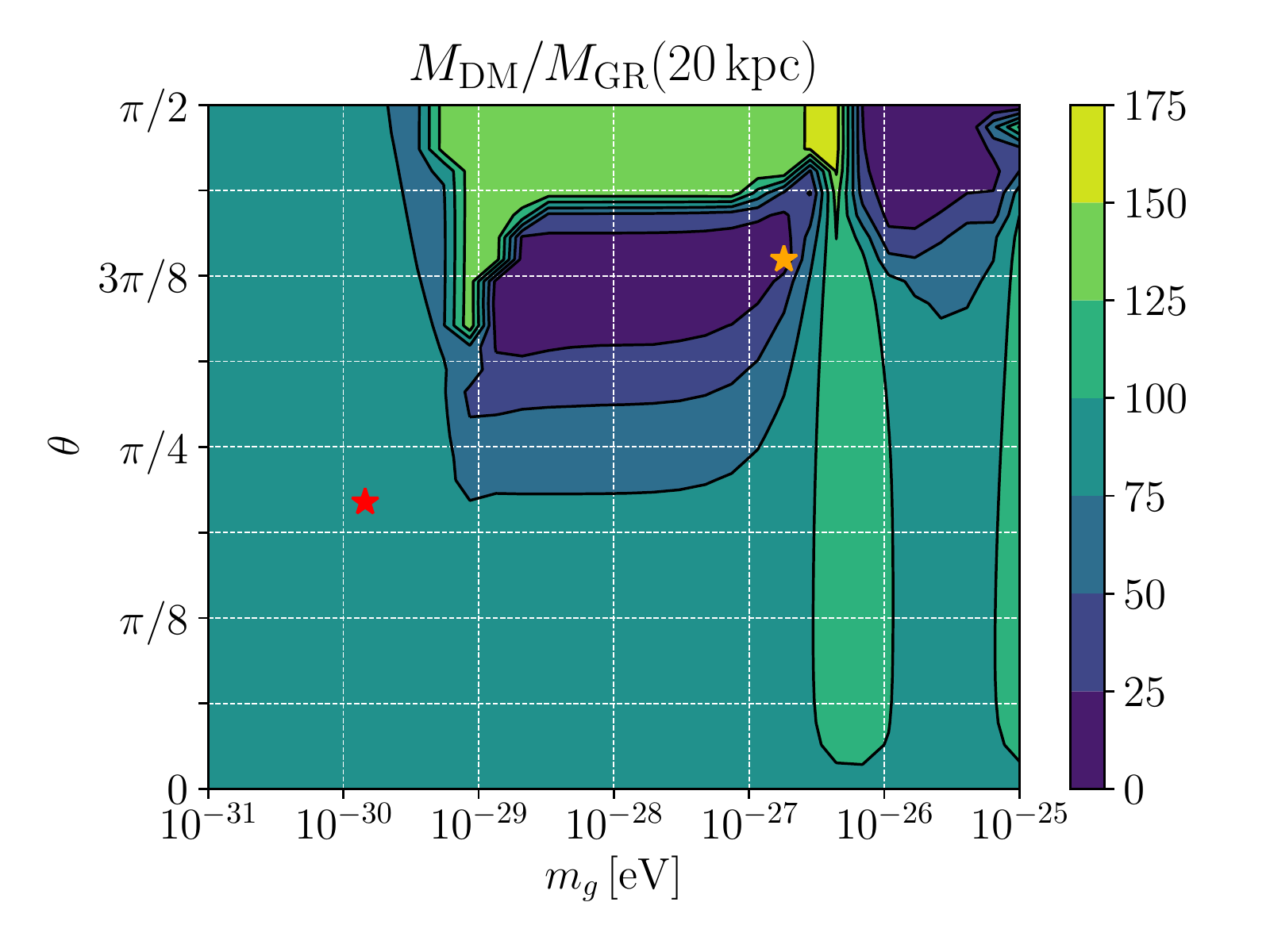}\label{fig:Massratio}}
	\caption{\label{fig:rotCurveScan}Parameter scan over the bigravity parameters $(m_g,\theta, \Delta r)$. The best fit point is marked in red, while the parameter point with the least DM content is marked in orange.}
\end{figure}

In order to fit the rotation curves, we assume that the relation $v(r) = \sqrt{r \frac{\mathrm{d}\phi}{\mathrm{d} r}}$ holds, and modify the potential $\phi$, following the above reasoning, as
\begin{equation} \label{eq:potentialDefinition}
	\phi (\vec{r}) = - G_N \int \mathrm{d}^3 \vec{r}\,'\ \rho(\vec{r}\,') \left[ \frac{\alpha(\theta)}{ |\vec{r} - \vec{r}\,'|} + \frac{\beta(\theta) e^{- m_g |\vec{r} - \vec{r}\,'|}}{ |\vec{r} - \vec{r}\,'|} \right],
\end{equation}
where the mixing angle is replaced by an effective mixing angle, as defined in Eq.~\eqref{eq:Vainshtein}, to account for the Vainshtein mechanism. Since the potential is a solution to a linear, inhomogeneous Poisson equation, the squared angular velocities related to the different mass components simply add up linearly as is the case in GR. We assume that the galaxy is composed of gas, a disk of ordinary (baryonic) matter and a DM halo. The gas component is inferred from the HI emission and is modelled by assuming a proportionality of mass and luminosity and an exponentially decreasing profile,
\begin{equation} \label{eq:surfaceMassGas}
	\Sigma(r) = \Sigma_0 e^{-r/r_0} \quad \Rightarrow \quad M_\text{HI}(r) = L\, \Sigma_0 \times \int_0^r \mathrm{d}r'\, \Sigma(r') = L\, \Sigma_0 \left(r_0- (r + r_0) e^{-r/r_0}  \right),
\end{equation} 
where the mass-to-light ratio $L$ is a free parameter. Following Ref.~\cite{Hashim:2014mka}, we model the baryonic disk such that, given a $1/r$ potential, the effective mass distribution would be 
\begin{equation}
	M_D(r) = 0.5 M_D^0 (3.2 x)^3 (I_0 K_0 - I_1 K_1),
\end{equation}
where $x \equiv r/R_\text{opt}$ and $R_\text{opt} \equiv 3.2 r_0$, and evaluating the modified Bessel functions $I_{0,1}$ and $K_{0,1}$ at $1.6x$. In this expression, only  $M_D^0$ is a free parameter, while $r_0$ is fixed by the HI observation.

The final component is the DM halo, which we assume to have an NFW profile,
\begin{equation}
	M_\text{DM}(r) = M_\text{DM}^0 \left[ \log\left( 1 + \frac{r}{r_h}\right) - \frac{r}{r+r_h}\right],
\end{equation}
with the free parameters $M_\text{DM}^0$ and $r_h$. 
\begin{figure}[t]
	\centering
	\subfloat[GR, $\theta = 0$]{
		\includegraphics[width=.45\textwidth]{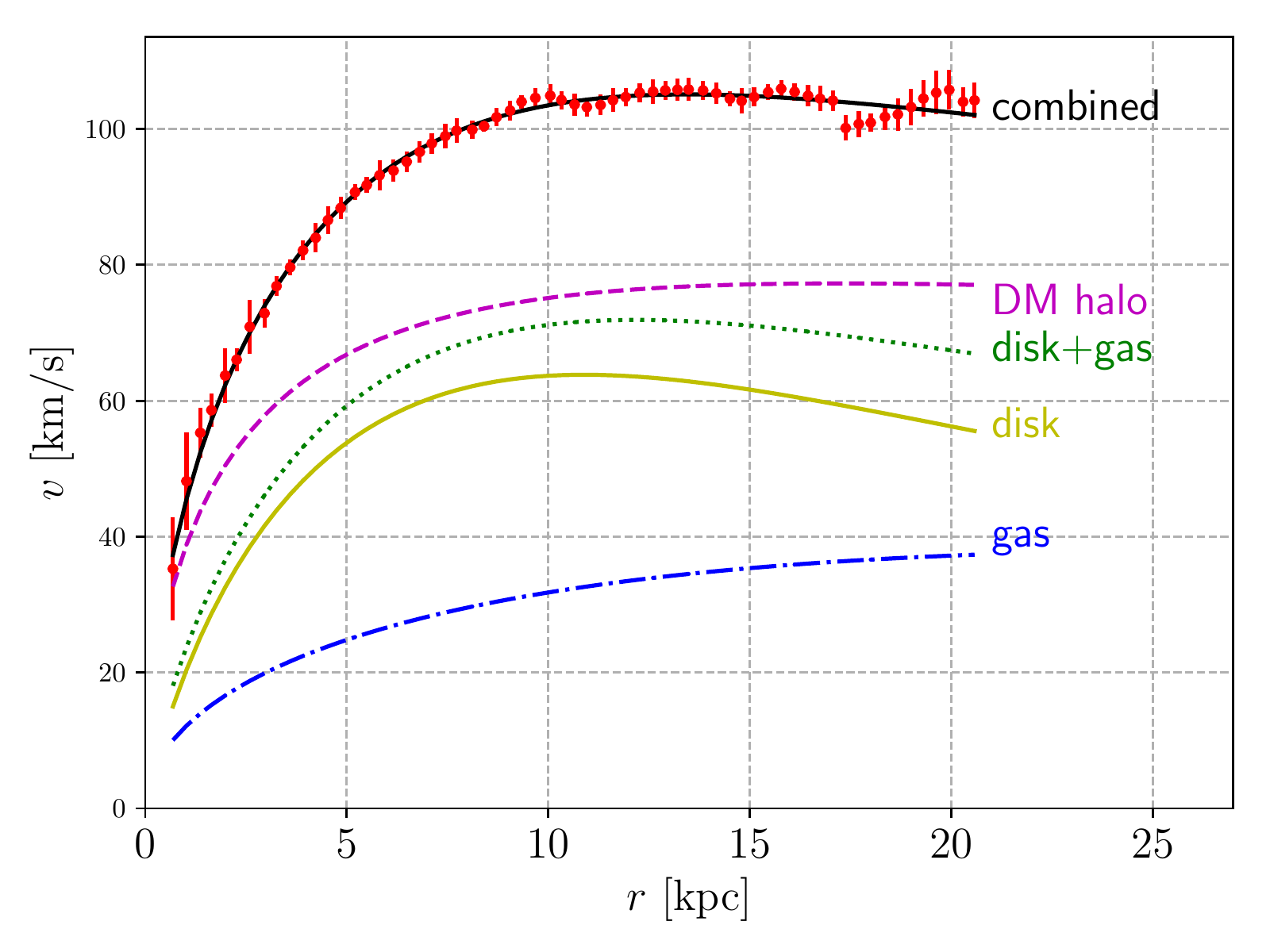}\label{fig:rotCurveFitGR}}
	\subfloat[$(m_g, \theta, \Delta r / r_V)=(1.4\cdot 10^{-30}\eV,  0.7, 0.5)$]	{
		\includegraphics[width=.45\textwidth]{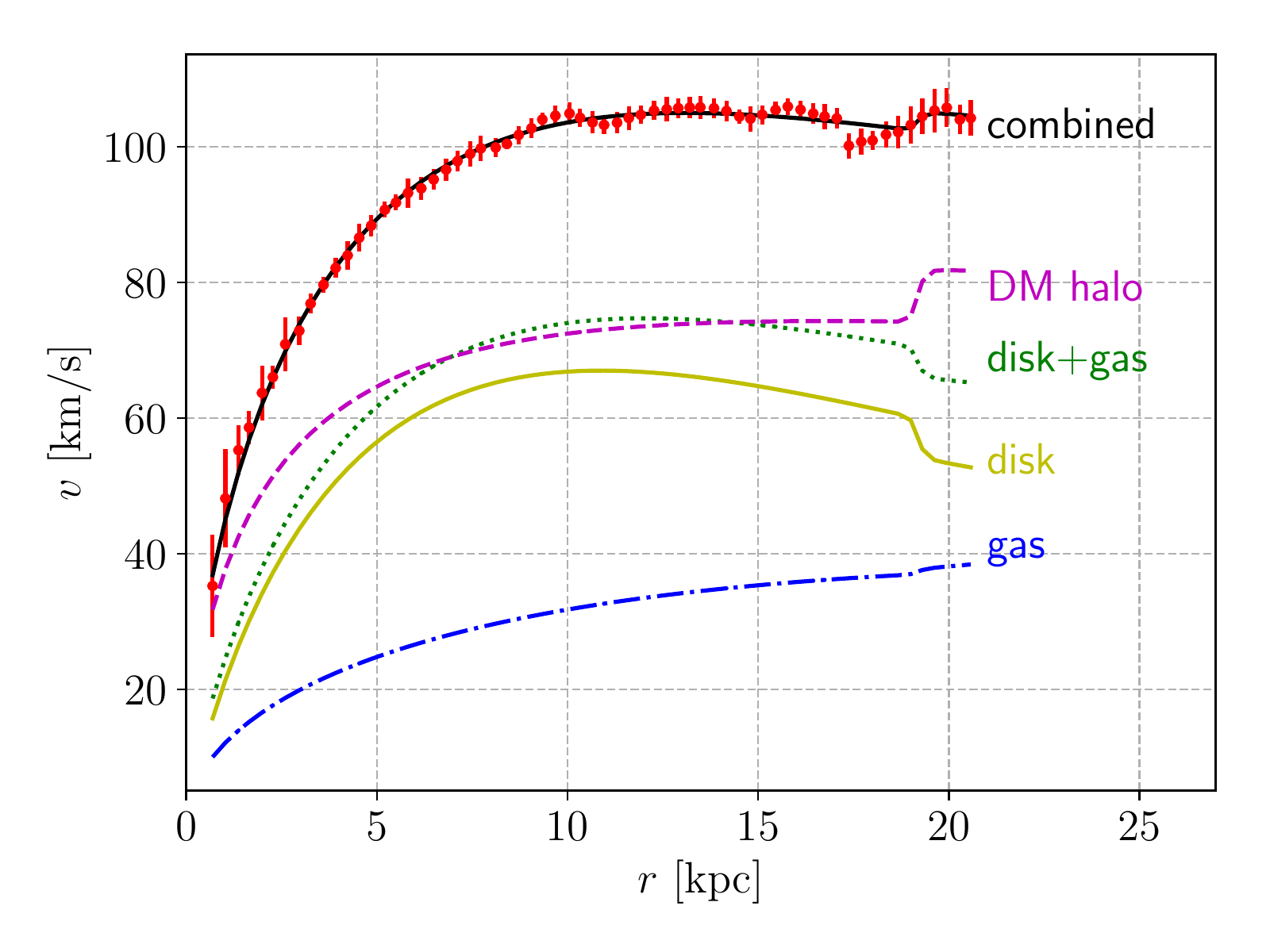}\label{fig:rotCurveFitBiGra}}
	\caption{\label{fig:rotCurveFit}ESO138-G014 rotation curve. Shown is the data with error bars, the final fit, and the three components making up the galaxy in our analysis. Panel (a) shows the fit in pure GR ($\chi_0^2=26$), while panel (b) shows the best-fitting point found in our analysis for which $\Delta \chi^2 \equiv (\chi^2-\chi_0^2)/\chi_0^2 = -0.16$.}
\end{figure}
Let us stress once again that it is not sufficient to assume $v^2(r) = r \frac{\mathrm{d}\phi}{\mathrm{d} r} = G_N M(r) / r$, because the Yukawa factor inside the integral in Eq.~\eqref{eq:potentialDefinition} no longer allows the contributions for $|\vec{r}\,'| > |\vec{r}|$ to cancel, as would be the case for the Newtonian relation. Due to computational limitations, we integrate the potentials analytically and implement these expressions in our numerical analysis. The rather lengthy expressions are shown in Appendix~\ref{app:potentials}.

The HI fit is straightforward and details are not shown here. In agreement with Ref.~\cite{Hashim:2014mka}, we find that $M_\text{HI} (20\,\mathrm{kpc}) = 6.4 \cdot 10^{9} M_\odot$. A fit to the rotation curve in GR is displayed in Fig.~\ref{fig:rotCurveFitGR} and illustrates the typical fall-off in velocity for the baryonic matter which is counteracted by the DM to fit the data. In GR, one needs approximately 57\% of the total mass inside the central galactic region to be DM, in order to fit the rotation curve.\footnote{We chose a \emph{fiducial} volume of radius $r=20\kpc$ to quantify how much DM is needed to fit the rotation curve. However, it should be emphasised that this does not correspond to the total mass of the DM halo, since the latter may extend far beyond the visible components. Alternatively, one may proceed as was done e.g.~in Ref.~\cite{deAlmeida:2018kwq}, where the mass parameter of the NFW density profile was used instead.}

We now repeat this fit with the modification indicated in Eq.~\eqref{eq:potentialDefinition}. For each value $(m_g, \theta)$, we obtain a fit to the data, similar to that shown in Fig.~\ref{fig:rotCurveFitGR}. The collection of these fits carried out over the bigravity parameter space is then summarised in Fig.~\ref{fig:rotCurveScan}, where we show the regions excluded by a $\chi^2$ analysis, the point representing the best fit (red star) and the point with lowest DM component (orange star) in panel~(a), while panel~(b) displays the ratio of DM mass and total mass within a `fiducial' volume of radius $20\kpc$.

Several important features of Fig.~\ref{fig:rotCurveScan} should be highlighted at this point. First, we observe that for large mixing angles and masses $\gtrsim 10^{-29}\eV$, more DM than in GR is needed. This is to be expected, as for large mixing we are close to massive GR, while the large mass induces an early exponential fall-off of the Yukawa force. Thus, the missing gravitational force has to be counteracted by even more DM than needed in GR.

The point in the $m_g$-$\theta$ plane with the lowest $\chi^2$, is marked with a red star in Fig.~\ref{fig:MassratioChisquared}.\footnote{While the region we display as exclusion region is based on a sound statistical foundation, we emphasise that the best-fit points shown in this work have a generally low statistical significance. We merely indicate these to highlight that the fits can be improved, in principle.} Notice that this point is within one order of magnitude of the best-fitting point shown for the lensing in Fig.~\ref{fig:ClusterExample}. While this point improves the value of $\chi^2$ compared to the GR result, it results in a DM component of only slightly lower mass, $M_\text{DM} / M_\text{GR} \simeq 95\%$, also cf.~Fig.~\ref{fig:rotCurveFitBiGra} where the corresponding fit to the rotation curves is shown. Indeed, we observe that most of the data points are inside the Vainshtein sphere, while the transition to the modified regime is used to fit a feature for the most out-lying data points.

\begin{figure}[t]
	\centering
	\subfloat[$(m_g, \theta, \Delta r/ r_V)=(2 \cdot10^{-27}\eV,  1.2, 0.04)$]	{
		\includegraphics[width=.45\textwidth]{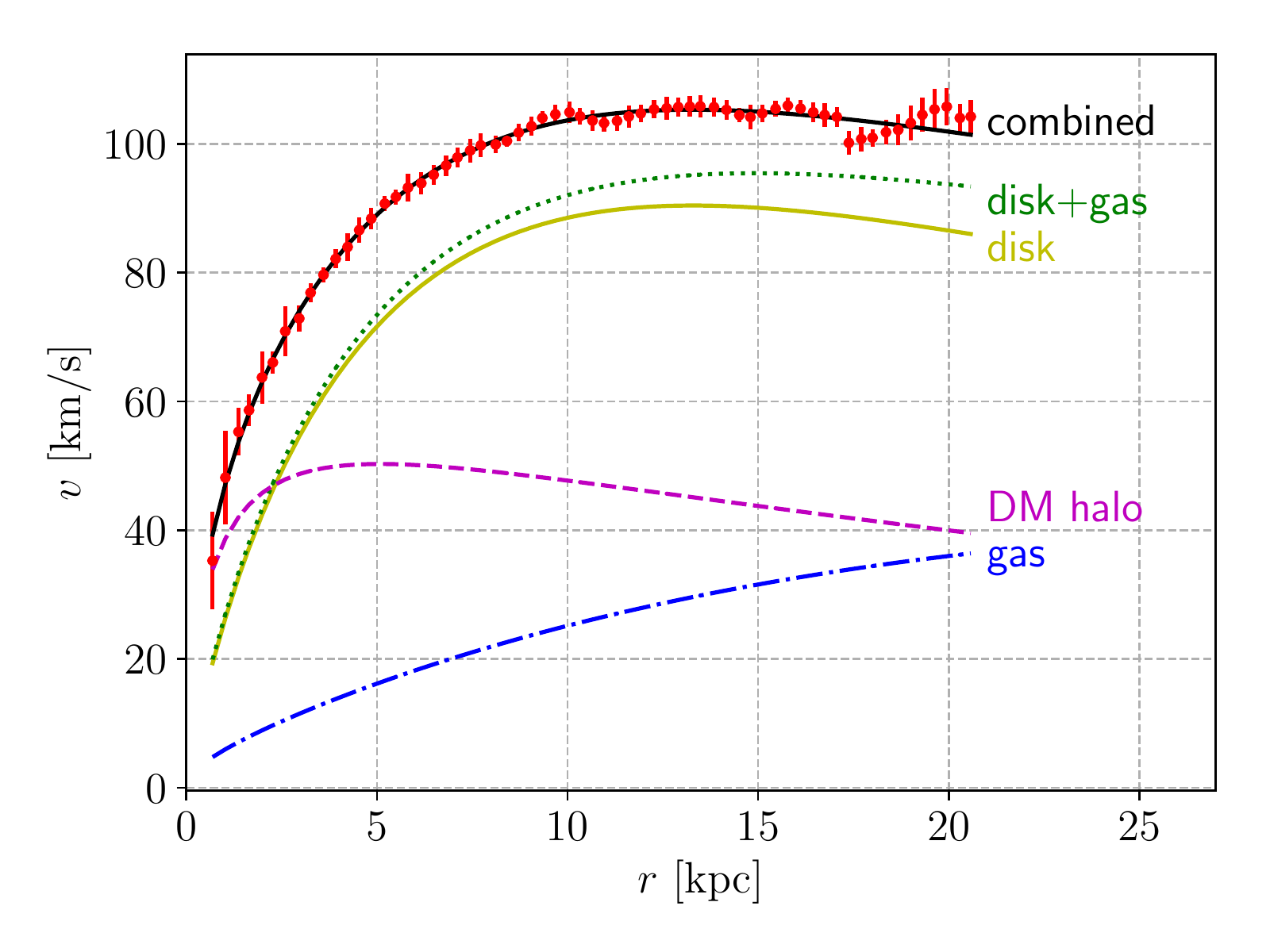}\label{fig:rotCurveminDM}}
	\subfloat[$(m_g, \theta, \Delta r/ r_V)=(2 \cdot 10^{-29}\eV, 1.0, 0.4)$]{
		\includegraphics[width=.45\textwidth]{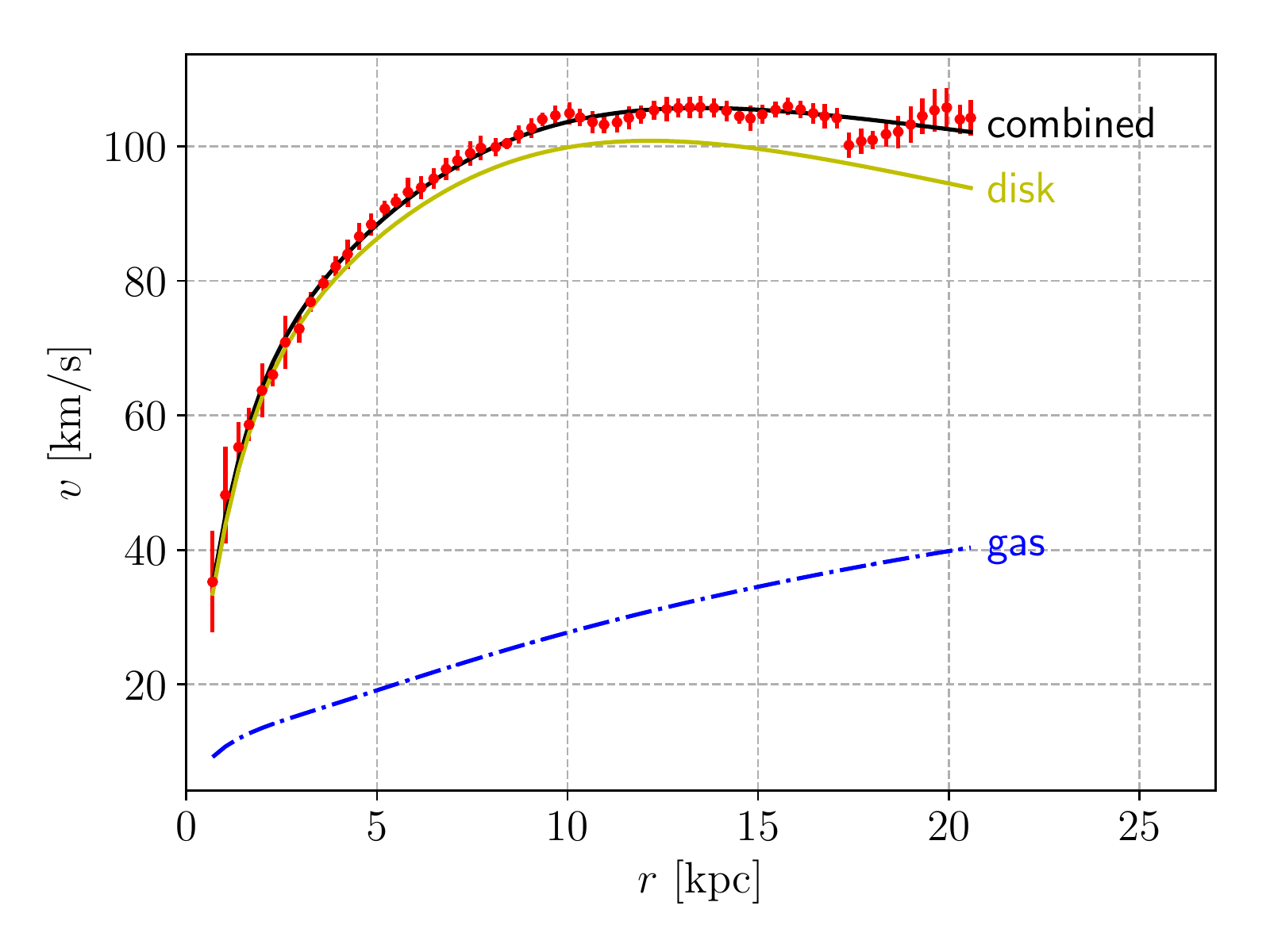}\label{fig:rotCurvepureBigra}}
	\caption{\label{fig:rotCurveFit2}Rotation curves fitted with low (left) and zero (right) DM mass fraction. We find $\Delta \chi^2$ of $-0.03$ and $+0.2$, respectively.}
\end{figure}

Next, one may observe that there is a region of intermediate mixing and masses $10^{-29} \eV \leq m_g \leq 10^{-27}\eV$, where the fit \emph{reduces} the amount of DM needed to a point where very little non-baryonic matter is required to obtain a good fit to the data. Again, we show an example in Fig.~\ref{fig:rotCurveminDM}, which corresponds to the point in parameter space which requires about a factor 15 less DM mass compared to GR, corresponding to $M_\text{DM} / M_\text{GR} \simeq 7\%$, and is thereby the point in parameter space with the smallest DM component. Finally, we point out to the reader that the Vainshtein mechanism sets in as we lower $m_g$ further, and no deviations from GR are expected.

A final exercise is trying to fit the data without any DM. This is shown in Fig.~\ref{fig:rotCurvepureBigra}, where we find the best-fitting point at a mass $m_g= 2 \cdot 10^{-29}\eV$, slightly below the best-fitting point found above. However, we note that the fit is slightly worse ($\Delta \chi^2 = 0.2$) when no DM halo is included, which disfavours this scenario.

The main conclusion of this section is that in most of the graviton parameter space where an effect on the rotation curve is expected, the degeneracy with the unknown DM component does not allow to place any limits. Only in a small range of graviton masses the $\chi^2$ function gets worse and disfavors this scenario. This graviton mass range, however, is already excluded by the cluster observations considered in the previous section. One caveat to this conclusion is the limit of pure massive gravity ($\theta=\pi/2$). For masses $m_g \gtrsim 10^{-26}\eV$ the Yukawa-suppression becomes too strong to yield any acceptable fit and this regime is therefore ruled out, cf.~Fig.~\ref{fig:combined} below. Any parameter point close to $\theta = \pi/2$ could then in principle be rendered viable at the expense of a very large DM component, counteracting the smallness of the long-range force. This behaviour has been confirmed with two more sets of galactic rotation curves containing low surface brightness galaxies \cite{KuziodeNaray:2007qi} and \cite{McGaugh:2001yc}, which we do not show explicitly at this point, but include in our combined exclusion plot which can be found in our discussion, Sec.~\ref{sec:discussion}. 

\subsection{A galaxy without Dark Matter}
Recently, astronomers have observed the galaxy NGC1052-DF2, which appears to contain no, or very little DM~\cite{vanDokkum:2018vup}. This interesting observation has been used to constrain the viability of modified gravitational theories such as modified Newtonian dynamics (MOND),~\cite{Milgrom:1983ca} which try to replace DM by modifying gravity on appropriate scales~\cite{Sanders:2002pf}. The absence of DM in an object of galactic size is therefore a challenging --~if not impossible~-- task for such theories. The point of view taken in this paper is quite different to the approach of MOND etc. Knowing that bigravity modifies the gravitational potential on a certain scale, we ask how much DM do we need to explain the observed data, instead of trying to replace it entirely. In this spirit, we can use the observation of the galaxy NGC1052-DF2, assuming the absence of DM in it, to put constraints of the viable parameter combinations in bigravity. 

Excluding the region $m_g > 10^{-25} \eV$ from our analysis, where one could add more and more DM to counteract the very strong Yukawa suppression, it is assumed that the galaxy contains no DM. Given the rotation curve data, we then find a fit to the data points and demand that the total mass remains below $3.2 \cdot 10^8 M_\odot$~\cite{vanDokkum:2018vup}. This excludes a small parameter region above $\theta > \frac{5\pi}{16}$ and $10^{-30} \eV \leq m_g \leq 10^{-27} \eV$. This region is included in our summary plot, Fig.~\ref{fig:combined}.

\section{Scaling behaviour}\label{sec:scaling}

\begin{figure}[t]
	\centering
	\includegraphics[width=.6\textwidth]{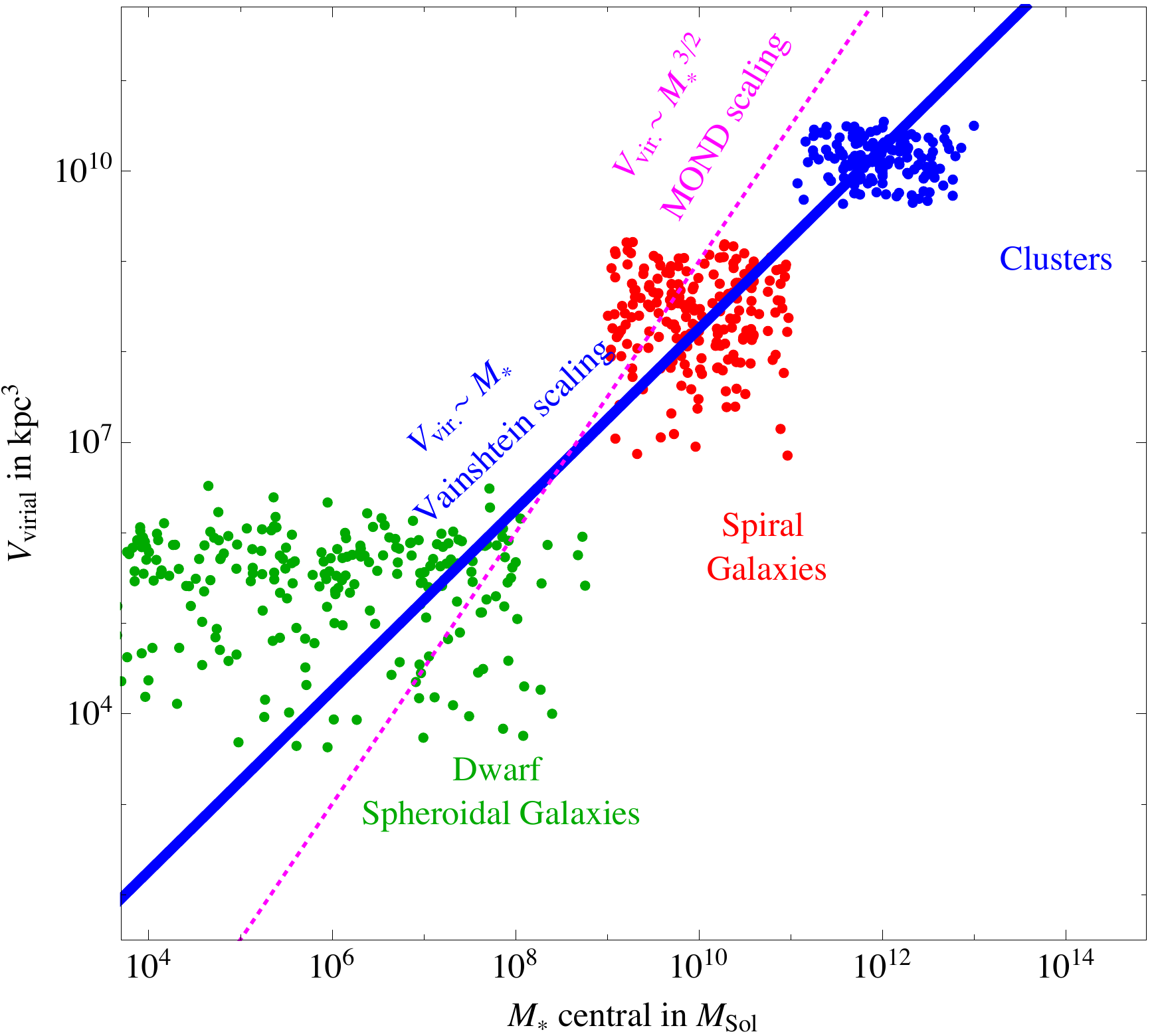}
	\caption{\label{fig:Scaling} The general scaling of the virial volume of gravitationally bound systems vs.~the estimated stellar mass component in the central region of the objects. The tentative scales of the objects are taken from \cite{Buckley:2017ijx}. The deviations from the GR behaviour set in at scales above the blue and magenta lines for theories with Vainshtein scaling and MOND, respectively.}
\end{figure}

In the previous sections we observed a curious effect. It turns out, that the induced metric in bigravity has quantitatively an effect of enhancing the gravitational effect of the matter source. This leads to emergent but spurious additional matter if the GR metric only is used. In particular, there are parameter choices at which the effect becomes relevant simultaneously at galactic and cluster scales.  We will now present a generic scaling argument based on the Vainshtein mechanism, wich supports this observation.

The gravitational anomaly, which is commonly explained entirely by the presence of a mainly gravitationally interacting, non-baryonic, non-relativistic matter component is observed in various physical systems. Thus, DM seems to have an effect on the gravitational interaction on vastly different scales.  Alternative explanations to the DM phenomenon involve theories with a new scale at which the gravitational law is modified. But how should this scale depend on the mass of the object to affect all astrophysical systems in which the DM effect is observed? To answer this question, we note that, if the critical length scale has a power law mass dependence $r_c \propto M^n$, then the critical volume around which gravity is modified scales as $V_c \propto M^{3n}$.

We compare in Fig.~\ref{fig:Scaling} the scaling behaviour of MOND and the scaling of the critical length scale found in massive extensions of gravity with data. It is widely known that while MOND can be adjusted to explain some of the DM effects on galaxy scales,~\cite{Sanders:2002pf,Gentile:2010xt,McGaugh:2016leg,Lelli:2017vgz,Li:2018tdo} it seems to fail in galaxy clusters~\cite{Sanders:2002ue,Angus:2006qy,Angus:2006ev}.\footnote{In fact, it was recently found in Ref.~\cite{Frandsen:2018ftj,Rodrigues:2018duc} that it is not possible to explain all galactic dynamics in MOND, and hence certain classes of MOND are disfavoured.}  The failure of MOND in clusters can be easily seen from Fig. (\ref{fig:Scaling}) as the scaling behaviour of the critical length scale in MOND is $r_c \propto M^{1/2}$, since the potential is modified at critical accelerations $a_0 = F/m \propto M r_c^{-2}$, and thus $V_c \propto M^{3/2}$. We observe that the virial volume of clusters lies below the dashed magenta MOND line in Fig. (\ref{fig:Scaling}), but only systems larger than the critical value indicated by this line can exhibit MOND-like modifications of gravity. Qualitatively, this is in agreement with the considerations of Ref.~\cite{Capozziello:2017rvz} that go beyond the non-relativistic MOND framework.

On the contrary the Vainshtein mechanism in massive gravity extensions has a critical length which scales as $r_V \propto ( M \lambda_g^2)^{1/3}$, with $\lambda_g$ being the wavelength of the graviton. Thus, the volume around which gravity is modified scales as the central mass source of the system $V_c \propto r_V^3 \propto M$. This is also trivially the case in DM halo models. Hence, we observe that massive gravity could alter our predictions of the amount of DM in compact gravitationally bound systems. A further unavoidable consequence of this modification of gravity is that the scale of the enhanced gravitational potential is linked to the baryonic matter scale. Observational evidence of this has been recently discussed in \cite{0004-637X-836-2-152}.

In general the Vainshtein mechanism in non-linear theories of massive gravity can be described by studying the equation of motion for the helicity-0 mode in the decoupling limit. As discussed in Ref.~\cite{Babichev:2013usa} we have the following equations of motion for the canonically normalised helicity-0 and helicity-2 modes:
\begin{equation}
\mathcal{E}_{\mu \nu}^{\alpha \beta} \hat{h}_{\alpha \beta} = \frac{T_{\mu \nu}}{M_{Pl}}\,, \quad
 3 \Box \phi +\mathcal{E}_{\phi}= \frac{T}{M_{Pl}}\,,
\end{equation}
 where $\mathcal{E}_{\mu \nu}^{\alpha \beta}$ follows from the linearisation of the Einstein tensor. When the term $\mathcal{E}_\phi$ is sub-dominant, it can be estimated that $\phi \sim T \equiv {T^\mu}_\mu$ and $\hat{h}_{\mu\nu} \sim T_{\mu\nu}$. Since the physical metric is a linear combination, $h_{\mu\nu} \sim \hat{h}_{\mu\nu} -\eta_{\mu_\nu}\phi$, this corresponds to the regime of massive gravity. Contrarily, when $\mathcal{E}_\phi$ dominates the evolution of $\phi$, the latter becomes non-dynamical and sub-dominant, $\phi \ll \hat{h}_{\mu\nu}$. Therefore the physical metric comprises only two degrees of freedom, $h_{\mu\nu} \sim \hat{h}_{\mu\nu}$, i.e.~GR is restored. This simple argument has been studied and confirmed in more detail in Refs.~\cite{Babichev:2009us,Babichev:2009ee,Babichev:2009jt,Babichev:2010jd}, and was also confirmed via numerical studies for massive and bimetric gravity~\cite{Gruzinov:2011mm,Volkov:2012wp}. 

Quite generally, the non-linear interactions lead to 
\begin{equation}
\mathcal{E}_\phi \propto \partial^{n-k+3} \phi^k/m_g^{n-1} M_{Pl}  = \partial^{n-k+3} \phi^k/\Lambda^n\,,
\end{equation}
where $\Lambda_n^n = M_{Pl}\, m_g^{n-1}$ is the strong coupling scale and leads to a Vainshtein radius $r_V = (M_{Pl}\, r_S)^{\frac{k-1}{n}} \Lambda^{-1}$. Our observation shows that only those theories will be relevant for DM observations on multiple scales in which $(k-1)/n = 1/3$. This implies that $\mathcal{E}_\phi \propto (\partial^{2} \phi)^k/ m_g^{3k-4} M_{Pl} $ and $r_V^3 =2 M^* M_{Pl}^{\frac{k}{1-k}} \lambda_g^{\frac{3k-4}{k-1}}$, with $M^*$ being the central mass. Thus, the restriction on the space of viable theories is that non-linear interactions of the helicity-0 modes have to only contain interactions with two powers of derivatives per one power of the field. Bigravity is an example of such a theory with $k=2$, but certainly not the only one.
 
\section{Discussion and Conclusion}
\label{sec:discussion}

Massive gravity, i.e.~gravity mediated exclusively by a massive spin-2 field, is subject to very strong constraints, most notably from weak lensing surveys, $m_g < 6 \cdot 10^{-32}\eV$~\cite{deRham:2016nuf,Desai:2017dwg,Rana:2018vxn}. Although this bound is somewhat model-dependent, it tightly restricts the available parameter space of the massive spin-2 mode. In this work, we have considered a framework where the second tensor field, which is a necessary ingredient for the graviton to become massive, becomes dynamical itself. Thereby, two tensor modes are present in the particle spectrum and mediate different forces, one long-range Coulomb-type force and one Yukawa-type force. A peculiar feature of massive spin-2 fields is the Vainshtein mechanism, an intrinsically non-linear effect, which is hard to implement in simple calculations. We have taken a pragmatic point of view, assuming that GR is continuously restored within a spherical volume of radius $r_V$, and studied the effects of the superimposed spin-2 mediated forces.

\begin{figure}[t]
\centering
\includegraphics[width=.8\textwidth]{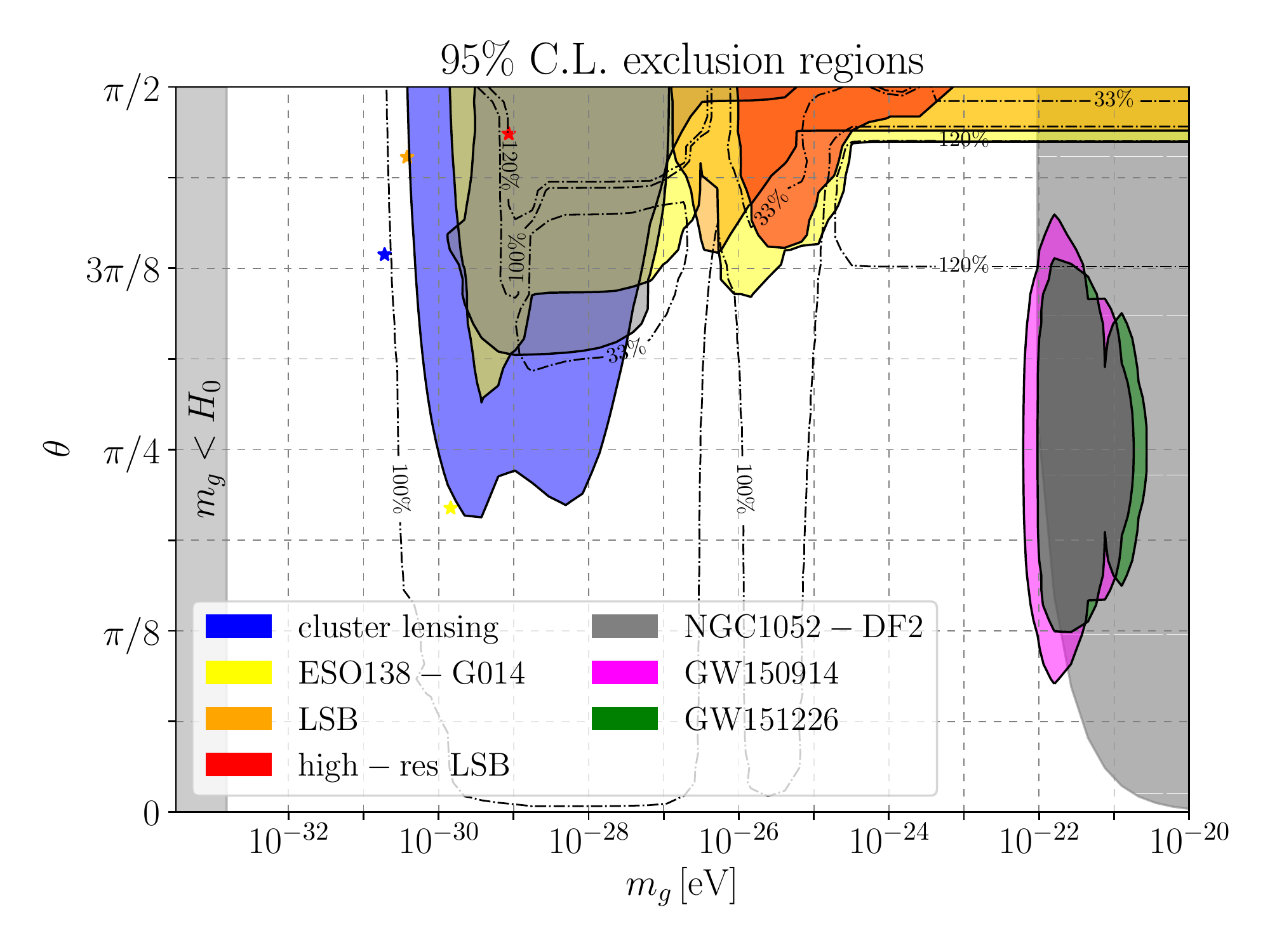}
\caption{\label{fig:combined}Combined 90\% exclusion regions derived in this and previous publications. The largest region is excluded due to the lensing constraints discussed in Sec.~\ref{sec:MACSlensing} (blue region). For completeness, we have also included the best-fitting points in corresponding color coding. The grey, hatched region is excluded from solar system tests~\cite{deRham:2014zqa}, while the magenta and green regions are excluded from gravitational wave observations~\cite{Max:2017flc}. Finally, we exclude the region where the graviton Compton wavelength exceeds the observable universe, $m_g < H_0$. The dashed contours indicate the predicted local DM density in percent of the GR expectation.}
\end{figure}

First, we have derived the deflection angle of light in bigravity, which is considerably more complex than in GR, and found a sufficiently accurate analytical approximation. To the best of our knowledge this is a new result and has not been derived in the literature so far. We have then used this result to study the differences expected in mass estimates of galaxy clusters inferred from weak lensing and hydrostatic considerations. We have found that the results may differ drastically from the GR prediction, where the masses are predicted to be equal, and which has allowed us to derive new constraints on the bigravity parameter space. We have also used this finding to explain a tentative anomaly in cluster mass ratios that has been observed and quantified with a `hydrostatic bias parameter' by the Planck collaboration~\cite{Ade:2015fva}, and that, depending on the analysis method, seems to indicate a deviation from the GR prediction. Furthermore, we have studied the MACS J1206.2-0847 cluster to illustrate how the data can be fitted by means of the modified mass ratio and derived constraints on the bigravity parameters $(m_g,\theta)$ wherever possible.

In a next step, we have used data from spiral galaxies to fit the corresponding rotation curves. We find that this, too, provides strong constraints, allows one to lower the required DM fraction in the galactic center to maximally 7\% of the amount needed in GR, or improve the fit to the rotation curve data; albeit, not simultaneously.

Fig.~\ref{fig:combined} shows a compilation of all constraints and best-fitting points found in our analysis. We have also included the bound from solar system tests (grey hatched region), and gravitational wave observations (green and magenta regions), see e.g.~\cite{deRham:2016nuf,Max:2017flc}. Finally, we disregard the region, where the graviton Compton wave length exceeds the observable Universe, $m_g < H_0$.

While this manuscript is certainly only a first step towards fully understanding the long-distance behaviour of bigravity, the results are quite encouraging to extend these surveys beyond our simple treatment, and, e.g., fully take into account non-linear effects and/or to study, in addition, the CMB, structure formation to name only a few more observables of interest. Given the scaling behaviour of the Vainshtein mechanism discussed in Sec.~\ref{sec:scaling}, studying the modified astrophysical and cosmological predictions is an important and powerful tool, since any best-fitting point for some observable, say for the tentative cluster anomaly, will have consequences for a variety of length scales and must therefore pass a series of constraints on these scales to be a viable explanation. Similarly, this scaling allows one to probe all of the available parameter space through whichever observation is most sensitive. Ultimately, it is crucial to further constrain this scenario, because of the fact that the local dark matter density and the gravity modification by theories with Vainshtein scaling have a considerable degeneracy. Otherwise, our searches for particle dark matter might be based on incorrect premises about the local density. 

\acknowledgments
J.S. would like to thank Annika Peter and Mads Frandsen for helpful discussions. J.S. is grateful to the support by the CP$^3$-Origins centre. The CP$^3$-Origins centre is partially funded by the Danish National Research Foundation, grant number DNRF90. M.P.~receives funding from the International Max Planck Research School for Precision Tests of Fundamental Symmetries (IMPRS-PTFS) and is enrolled at the University of Heidelberg. S.M.~acknowledges support by the DFG grant BA 1369 / 20-2 . 

\appendix
   
\section{Gravitational potential and tangential velocity in bigravity} \label{app:potentials}
In this appendix, we present the results of the integration of the gravitational source equation,
\begin{equation}
	\phi (\vec{r}) = - G_N \int \mathrm{d}^3 \vec{r}\,'\ \rho(\vec{r}\,') \left[ \frac{\alpha(\theta)}{ |\vec{r} - \vec{r}\,'|} + \frac{\beta(\theta) e^{- m_g |\vec{r} - \vec{r}\,'|}}{ |\vec{r} - \vec{r}\,'|} \right],
\end{equation}
for the gas, disk, and DM halo densities, respectively. We exploit the fact that in order to obtain the tangential velocity, $v(r) = \sqrt{r \frac{\mathrm{d}\phi}{\mathrm{d}r}}$, we can exchange the integration and differentiation in the above equation.

For the gas component, we find
\begin{equation}
	v^2_\text{gas} (r) =  G_N \left[  \alpha(\theta)  \frac{M_{\text{gas}}(r, r_0 , m_0) }{r}+ \beta(\theta) \frac{m_0}{2}   \frac{ (e^{-m_g \, r} \, (1+ m_g\,r) - e^{-r/r_0} \, (r+r_0)/r_0)}{ 1 - m_g^2 r_0^2} \right]\,, 
\end{equation}
where we assumed a surface mass density, as described in Eq.~\eqref{eq:surfaceMassGas}, and $m_0 = L \Sigma_0$. Similarly, assuming that the visible disk has a density $\rho(r) = M_D^0 / (r + r_0) / (r^2 + r_0^2)$, we find
\begin{align}
	&v^2_\text{disk} (r) =  G_N \left\lbrace  \alpha(\theta)  \frac{M_{\text{disk}}(r, r_0 , M_D^0) }{r} + \beta(\theta) \frac{M_D^0}{2}  \left[ \frac{1 + m_g \, r}{ x_0\, r}  e^{-m_g \, r} \Big( - \frac{e^{-x_0}}{2} \mathrm{Ei}(m_g\, (r+r_0))  \right.\right.  \nonumber\\
	& + \cos(x_0)\, \mathrm{Si} (x_0) - \sin(x_0)\, \mathrm{Ci}(x_0) + \cosh(x_0)\, \mathrm{Shi}(x_0)  - \sinh(x_0) \, \mathrm{Chi} (x_0) - \frac{\pi}{2} \sin(x_0)   \nonumber\\
	& +  \cos(x_0)\, \Re\left\lbrace \left(\frac{1}{2}+\frac{i}{2}\right) \,  \mathrm{Ei}(m_g \, (r+i r_0))\right\rbrace + \sin(x_0)\, \Re\left\lbrace\left(\frac{1}{2}+\frac{i}{2}\right) \, \mathrm{Ei}(m_g\,(r-i r_0))\right\rbrace\Big)  \nonumber\\
	& + \frac{1 - m_g\,r}{ 2\,x_0\, r} e^{m_g \, r} \Big( \pi [\cos(x_0) + \sin(x_0)] - \Re\left\lbrace \left(\frac{1}{2}+\frac{i}{2}\right) \,  \mathrm{Ei}(-m_g \, (r+i r_0))\right\rbrace  \nonumber \\
	& + \left.\left.  \sin(x_0)\, \Re\left\lbrace\left(\frac{1}{2}+\frac{i}{2}\right) \, \mathrm{Ei}(- m_g\,(r-i r_0))\right\rbrace + e^{x_0} \, \mathrm{Ei}(-m_g\,(r+r_0)) \Big)  \right]\right\rbrace\,,
\end{align}
where we have used the short-hand notation $x_0 \equiv m_g \, r_0$, and the exponential integral function is defined as
\begin{equation}
	\mathrm{Ei} (x) = - \int_{-x}^\infty \mathrm{d}t\, \frac{e^{-t}}{t} \,,
\end{equation}
with appropriate linear combinations defining the trigonometric [$\mathrm{Ci}(x)$, $\mathrm{Si}(x)$] and hyperbolic integral functions  [$\mathrm{Chi}(x)$, $\mathrm{Shi}(x)$].

And finally, for the DM halo we assume the NFW profile, $\rho_\text{NFW} =  M_\text{DM}^0 / r / (r+r_h)^2$,
\begin{equation}
\begin{split}
	v^2_\text{NFW} (r) =  \frac{G_N}{r} \Bigg\lbrace  \alpha(\theta)  M_{\text{halo}}(r, r_h , M_\text{DM}^0) + \beta(\theta) \frac{M_\text{DM}^0}{2}  \Bigg[ \frac{2/r}{r + r_h} - ( 1 - m_g \, r) \, e^{m_g\, (r + r_h)} \times\\
	 \times \mathrm{Ei}(-m_g (r + r_h)) + (1 + m_g\, r)  \big( e^{-m_g\, (r - r_h)} \, \mathrm{Ei}(-m_g r_h)  +  e^{-m_g\, (r + r_h)}\times \\ 
	 \times\left[ \mathrm{Ei}(m_g\,r_h) - \mathrm{Ei}(m_g\,(r + r_h)) \right]\big) \Bigg]\Bigg\rbrace.
\end{split}
\end{equation}
This result is consistent with known results  in the literature~\cite{deAlmeida:2018kwq}.

\bibliographystyle{apsrev4-1}
\bibliography{literature}   
   
\end{document}